\documentclass{jfm}
\usepackage{graphicx}
\usepackage{epstopdf, epsfig}

\usepackage{amstext}
\usepackage{amsmath}
\usepackage{amsfonts,amssymb}
\usepackage{framed}
\usepackage{algorithm}
\usepackage{algorithmicx}
\usepackage{algcompatible}
\usepackage{subcaption}
\usepackage{lipsum}
\usepackage{bm}
\usepackage{xcolor}
\usepackage{pdfpages}

\def\dashL{\bm{\mbox{--~--~--}}}

\def\Lcirc{\mbox{---}~{\hspace*{-.12in}$\circ$}\hspace*{-.12in}~\mbox{---}}

\shorttitle{Phase-resolved ocean wave forecast with data assimilation}
\shortauthor{G. Wang and Y. Pan}

\title{Phase-resolved ocean wave forecast with ensemble-based data assimilation}

\author{Guangyao Wang\aff{1}
  \and Yulin Pan\aff{1}\corresp{\email{yulinpan@umich.edu}} }

\affiliation{\aff{1}Department of Naval Architecture and Marine Engineering, University of Michigan,
Ann Arbor, MI 48109, USA}

\begin{document}

\maketitle

\begin{abstract}
Through ensemble-based data assimilation (DA), we address one of the most notorious difficulties in phase-resolved ocean wave forecast, regarding the deviation of numerical solution from the true surface elevation due to the chaotic nature of and underrepresented physics in the nonlinear wave models. In particular, we develop a coupled approach of the high-order spectral (HOS) method with the ensemble Kalman filter (EnKF), through which the measurement data can be incorporated into the simulation to improve the forecast performance. A unique feature in this coupling is the mismatch between the predictable zone and measurement region, which is accounted for through a special algorithm to modify the analysis equation in EnKF. We test the performance of the new EnKF-HOS method using both synthetic data and real radar measurements. For both cases (though differing in details), it is shown that the new method achieves much higher accuracy than the HOS-only method, and can retain the phase information of an irregular wave field for arbitrarily long forecast time with sequentially assimilated data.
\end{abstract}

\begin{keywords}
Authors should not enter keywords on the manuscript, as these must be chosen by the author during the online submission process and will then be added during the typesetting process (see http://journals.cambridge.org/data/\linebreak[3]relatedlink/jfm-\linebreak[3]keywords.pdf for the full list)
\end{keywords}

\section{Introduction}
Accurate prediction of ocean waves plays a significant role in the industries of shipping, oil \& gas, aquaculture, ocean renewable energy, coastal and offshore construction. In the past few decades, both phase-averaged and phase-resolved wave models have been developed. The phase-averaged wave models, which provide statistical descriptions in terms of the wave spectrum, have been widely used in the operational forecast of global and regional sea states~\citep{booij1999third,tolman2009user}. Despite their wide applications and success, phase-averaged models have limitations of providing no information on the individual deterministic waves. For example, rogue waves, which often appear sporadically and potentially cause enormous damages to offshore structures and ships~\citep{broad2006rogue,nikolkina2011rogue}, cannot be captured. On the other hand, phase-resolved models can forecast the evolution of individual waves, but receive much less attention historically, partly due to the difficulty in obtaining the phase-resolved ocean surface as initial conditions. This has now been largely ameliorated with the recent development of sensing technologies and wave field reconstruction algorithms~\citep[e.g.][]{reichert2004x,nouguier2013nonlinear,gallego2011variational,nwogu2010surface,lyzenga2015real,qi2016phase,qi2018nonlinear,desmars2018phase}. For example, the Doppler coherent marine radars have been applied to measure the radial surface velocity field, based on which the field of both velocity potential and surface elevation can be reconstructed in real time~\citep{nwogu2010surface,lyzenga2015real}.

Given the reconstructed surface elevation and velocity potential as initial conditions, the evolution of the wave field can be predicted by linear or nonlinear phase-resolved wave models. Although the linear models yield low computational cost, their prediction horizon is severely limited~\citep[e.g.][]{qi2017phase,blondel2010experimental}. For nonlinear models, the Euler equations governing the free surface need to be numerically integrated. One efficient numerical algorithm to achieve this goal, based on high-order spectral (HOS) method, is developed by~\cite{dommermuth1987high};~\cite{west1987new}, with later variants such as~\cite{craig1993numerical};~\cite{xu2009numerical}. The novelties of these algorithms lie in the development of an efficient spectral solution of a boundary value problem involved in the nonlinear wave equations, which is neglected in the linear wave models with the sacrifice of accuracy. In recent years, HOS has been developed for short-time predictions of large ocean surface taking radar measurements as initial conditions~\citep{xiao2013study}. However, due to the significant uncertainties involved in the realistic forecast (e.g., imperfect initial free surface due to measurement and reconstruction errors; the effects of wind, current, etc., that are not accurately accounted for) as well as the chaotic nature of the nonlinear evolution equations, the simulation may deviate quickly from the true wave dynamics~\citep{annenkov2001predictability}. Because of this critical difficulty, operational phase-resolved wave forecast has been considered as a ``hopeless adventure'' to pursue~\citep{janssen2008progress}.  

The purpose of this paper, however, is to show that the dilemma faced by the phase-resolved wave forecast can be largely addressed by data assimilation (DA), i.e., a technique to link the model to reality by updating the model state with measurement data~\citep{evensen2003ensemble,evensen2009data,bannister2017review}. Mathematically, the principle of DA is to minimize the error of analysis (i.e., results after combining model and measurements), or in a Bayesian framework, to minimize the variance of the state posterior given the measurements~\citep{evensen1994sequential,evensen2003ensemble,carrassi2018data}. Depending on formulations and purposes, two categories of DA algorithms exist, namely the variational-based and the Kalman-filter-based approaches. Among the limited studies to couple DA with phase-resolved wave models, most use the variational-based method, where the purpose is to find the optimal initial condition to minimize a cost function measuring the distance between the model prediction and data in future times~\citep{aragh2008variation, fujimoto2020ensemble, qi2018nonlinear}. These methods, however, are not directly applicable to operational forecast due to their requirement of future data far after the analysis state (in contrast to the realistic situation where data becomes available sequentially in time). On the other hand, the Kalman-filter-based approach allows data to be sequentially assimilated, by updating the present state as a weighted average of prediction and data according to the error statistics. The only attempt (based on the authors' knowledge) to couple such an approach with phase-resolved wave model is~\cite{yoon2015explicit}, which however assumes linear propagation of the model error covariance matrix, thus limiting its application only to wave fields of small steepness. More robust methods based on ensemble Kalman filter (EnKF, i.e., with error statistics estimated by ensemble of model simulations), which lead to many recent successes in geosciences~\citep{carrassi2018data}, have never been applied to phase-resolved wave forecast. Moreover, most existing work, if not all, use only synthetic data for the validation of their methods, which ignores the realistic complexity that should be incorporated into the forecast framework, such as the mismatch between the predictable zone and measurement region, and the under-represented physics in the model.

In the present work, we develop the sequential DA capability for nonlinear wave models, by coupling ensemble Kalman filter (EnKF) with HOS. The coupling is implemented in a straightforward manner due to the non-intrusive nature of EnKF, i.e., the HOS code can be directly reused without modification~\citep{evensen2003ensemble,evensen2009data}. The new EnKF-HOS solver is able to handle long-term forecast of the ocean surface ensuring minimized analysis error by combining model prediction and measurement data. The possible mismatch of the predictable zone (which shrinks in time) and measurement region (which moves with, say, ship speed) is accounted for by a new analysis equation in EnKF. To improve the robustness of the algorithm (i.e., address other practical issues such as misrepresentation of the error covariance matrix due to finize ensemble size and underrepresented physics in the model), we apply both adaptive covariance inflation and localization, which are techniques developed elsewhere in the EnKF community~\citep{anderson1999monte,anderson2007adaptive,hamill2005accounting,carrassi2018data}. We test the performance of the EnKF-HOS method against synthetic and realistic radar data, which shows consistent and significant improvement in forecast accuracy over the HOS-only method in both cases. For the former, we further characterize the effect of parameters in EnKF on the performance. For the latter, we show that the EnKF-HOS method can retain the wave phases for arbitrarily long forecast time, in contrast to the HOS-only method which loses all phase information in a short time.

The paper is organized as follows. The problem statement and detailed algorithm of EnKF-HOS method are introduced in $\S$\ref{sec:PF}. The validation and benchmark of the method against synthetic and realistic radar data are presented in $\S$\ref{sec:numres}. We give a conclusion of the work in $\S$\ref{sec:conc}.

\section{Mathematical formulation and methodology}\label{sec:PF}

\subsection{Problem statement}
We consider a sequence of measurements of the ocean surface in spatial regions $\mathcal{M}_j$, with $j=0,1,2,3, \cdots$ the index of time $t$. In general, we allow $\mathcal{M}_j$ to be different for different $j$, reflecting a mobile system of measurement, e.g, a shipborne marine radar or moving probes. We denote the surface elevation and surface potential, reconstructed from the measurements in $\mathcal{M}_j$, as $\eta_{\text{m},j}(\boldsymbol{x})$ and $\psi_{\text{m},j}(\boldsymbol{x})$ with $\boldsymbol{x}$ the two-dimensional spatial coordinates, and assume that the error statistics associated with $\eta_{\text{m},j}(\boldsymbol{x})$ and $\psi_{\text{m},j}(\boldsymbol{x})$ is known a priori from the inherent properties of the measurement equipment. 

In addition to the measurements, we have available a wave model that is able to simulate the evolution of the ocean surface (in particular $\eta(\boldsymbol{x},t)$ and $\psi(\boldsymbol{x},t)$) given initial conditions. Our purpose is to incorporate measurements $\eta_{\text{m},j}(\boldsymbol{x})$ and $\psi_{\text{m},j}(\boldsymbol{x})$ into the model simulation sequentially (i.e., immediately as data become available in time) in an optimal way such that the analysis of the states $\eta_{\text{a},j}(\boldsymbol{x})$ and $\psi_{\text{a},j}(\boldsymbol{x})$ (thus the overall forecast) are most accurate.   

\subsection{The general EnKF-HOS coupled framework}
\label{sec:enkfhos}
\begin{figure}
  \centerline{\includegraphics[scale =0.5]{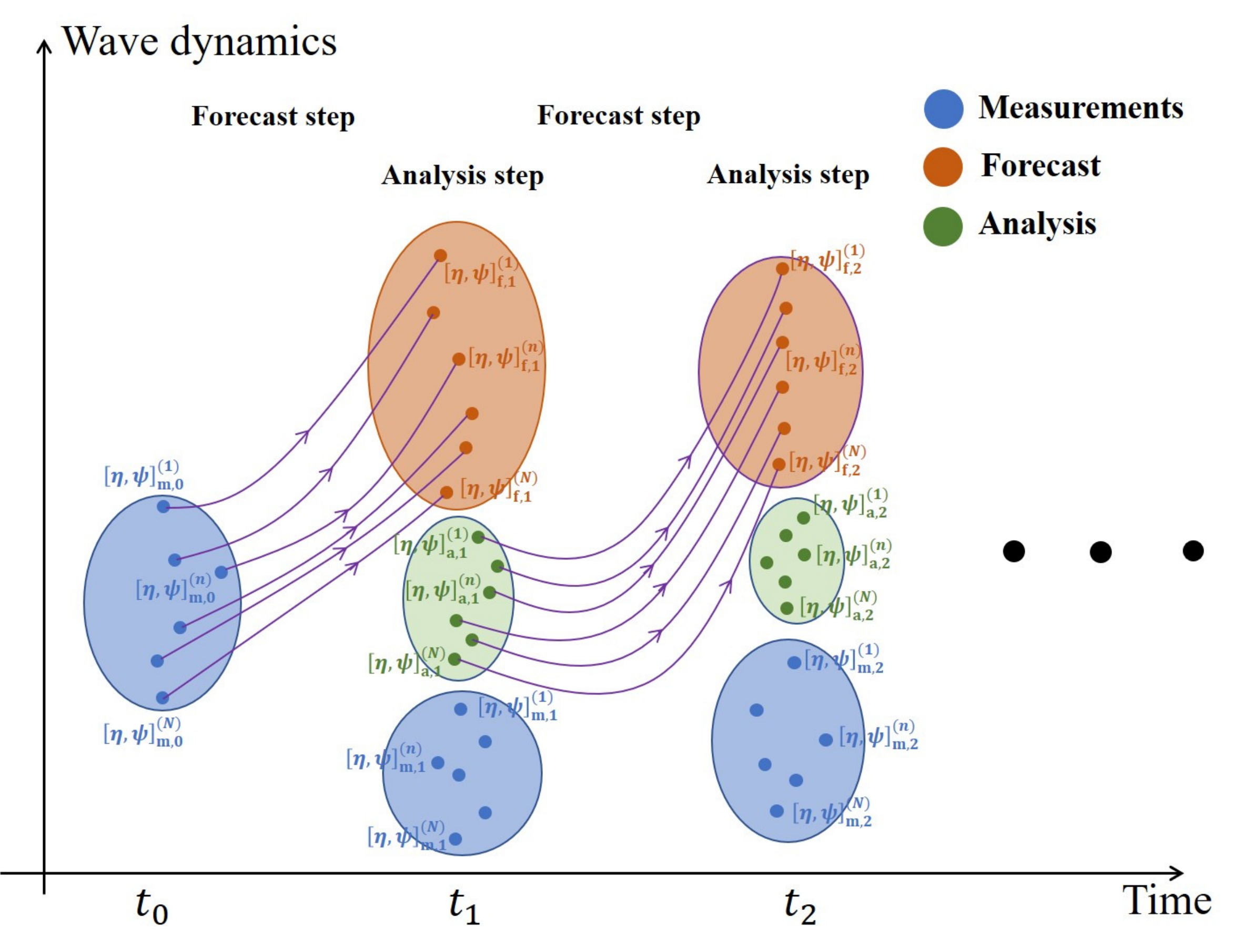}}
  \caption{Schematic illustration of the EnKF-HOS coupled framework. The size of ellipse represents the amount of uncertainty. We use short notations $[\eta,\psi]^{(n)}_{*,j}$ to represent $\eta^{(n)}_{*,j}(\boldsymbol{x}), \psi^{(n)}_{*,j}(\boldsymbol{x})$ with $*=\text{m},\text{f},\text{a}$ for measurement, forecast, and analysis, and $j=0,1,2$.}
\label{fig:enkfhos}
\end{figure}

In this study, we use HOS as the nonlinear phase-resolved wave model, coupled with the ensemble Kalman filter (EnKF) for data assimilation (DA). Figure~\ref{fig:enkfhos} shows a schematic illustration of the proposed EnKF-HOS coupled framework. At initial time $t=t_0$, measurements $\eta_{\text{m},0}(\boldsymbol{x})$ and $\psi_{\text{m},0}(\boldsymbol{x})$ are available, according to which we generate ensembles of perturbed measurements, $\eta_{\text{m},0}^{(n)}(\boldsymbol{x})$ and $\psi_{\text{m},0}^{(n)}(\boldsymbol{x})$, $n=1,2,...,N$, with $N$ the ensemble size, following the known measurement error statistics (see details in \S\ref{sec:pert}). A forecast step is then performed, in which an ensemble of $N$ HOS simulations are conducted, taking $\eta_{\text{m},0}^{(n)}(\boldsymbol{x})$ and $\psi_{\text{m},0}^{(n)}(\boldsymbol{x})$ as initial conditions for each ensemble member $n$ (\S\ref{sec:hos}), until $t=t_1$ when the next measurements become available. At $t=t_1$, an analysis step is performed where the model forecasts $\eta_{\text{f},1}^{(n)}(\boldsymbol{x})$ and $\psi_{\text{f},1}^{(n)}(\boldsymbol{x})$ are combined with new perturbed measurements $\eta_{\text{m},1}^{(n)}(\boldsymbol{x})$ and $\psi_{\text{m},1}^{(n)}(\boldsymbol{x})$ to generate the analysis results $\eta_{\text{a},1}^{(n)}(\boldsymbol{x})$ and $\psi_{\text{a},1}^{(n)}(\boldsymbol{x})$ (\S\ref{sec:enkf}). The analysis step ensures minimal uncertainty represented by the analysis ensembles (figure~\ref{fig:enkfhos}), which is mathematically accomplished through the EnKF algorithm. A new ensemble of HOS simulations are then performed taking $\eta_{\text{a},1}^{(n)}(\boldsymbol{x})$ and $\psi_{\text{a},1}^{(n)}(\boldsymbol{x})$ as initial conditions, and the procedures are repeated for $t=t_2, t_3, \cdots$ until the desired forecast time $t_{\text{max}}$ is reached. The details of each step are introduced next in the aforementioned sections, with the addition of inflation/localization algorithm to improve the robustness of EnKF included in \S 2.6, and treatment of the mismatch between predictable zone and measurement region by modifying the EnKF analysis equation in \S 2.7. The full process is finally summarized in \S 2.8 with  Algorithm~\ref{al:asimilation}.

\subsection{Generation of the ensemble of perturbed measurements}
\label{sec:pert}
As described in \S 2.2, ensembles of perturbed measurements are needed at $t=t_j$, as the initial conditions of $N$ HOS simulations for $j=0$, and the input of the analysis step for $j\geq 1$. We collect and denote these ensembles by

\begin{equation}
\boldsymbol{S}_{\text{m},j}=[s_{\text{m},j}^{(1)},s_{\text{m},j}^{(2)},\cdots s_{\text{m},j}^{(n)},\cdots s_{\text{m},j}^{(N-1)},s_{\text{m},j}^{(N)}]\in \mathbb{R}^{d_j \times N},
\label{eq:etaeng}
\end{equation}
where $s$ represents the state variables of surface elevation $\eta$ or surface potential $\psi$, and $\boldsymbol{S}$ the corresponding ensemble. This simplified notation will be used hereafter when necessary to avoid writing two separate equations for $\eta$ and $\psi$. $s_{\text{m},j}^{(n)}$ with $n=1,2,...,N$ is the $n_{th}$ member of the perturbed measurements. $d_j$ denotes the number of elements in the measurement state vector of either $\eta$ or $\psi$ at $t=t_j$. Without loss of generality, in this work, we use constant $d_j=d$ for $j
\geq1$, and choose $d_0$ for the convenience of specifying model initial condition (see details in \S3).

To generate each ensemble member $s_{\text{m},j}^{(n)}$ from measurements $s_{\text{m},j}$, we first produce $\eta_{\text{m},j}^{(n)}$ from
\begin{equation}
\eta_{\text{m},j}^{(n)}(\boldsymbol{x})={\eta}_{\text{m},j}(\boldsymbol{x})+w^{(n)}(\boldsymbol{x}),
\label{eq:etaini}
\end{equation}
where $w^{(n)}(\boldsymbol{x})$ is the random noise following a zero-mean Gaussian process with spatial correlation function~\citep{evensen2003ensemble,evensen2009data} 
\begin{equation}
C(w^{(n)}(\boldsymbol{x}_1),w^{(n)}(\boldsymbol{x}_2))=\begin{cases}c\exp\left(-\displaystyle\frac{|\boldsymbol{x}_1-\boldsymbol{x}_2|^2}{a^2}\right) &\text{for}~ |\boldsymbol{x}_1-\boldsymbol{x}_2|\leq\sqrt{3}a,\\
0~&\text{for}~|\boldsymbol{x}_1-\boldsymbol{x}_2|>\sqrt{3}a.
\end{cases}
\label{eq:noise1}
\end{equation} 
In~\eqref{eq:noise1}, $c$ is the variance of $w^{(n)}(\boldsymbol{x})$ and $a$ the de-correlation length scale, both of which practically depends on the characteristics of the measurement devices (and thus assumed known in priori). The perturbed measurement of surface potential $\psi_{\text{m},j}^{(n)}$ is reconstructed from $\eta_{\text{m},j}^{(n)}$ based on the linear wave theory,
\begin{equation}
 \psi_{\text{m},j}^{(n)}(\boldsymbol{x})=\int\frac{i\omega(\boldsymbol{k})}{|\boldsymbol{k}|}\tilde{\eta}_{\text{m},j}^{(n)}(\boldsymbol{k})e^{i\boldsymbol{k}\cdot\boldsymbol{x}}d\boldsymbol{k},
 \label{eq:lpsi}
\end{equation}
where $\tilde{\eta}^{(n)}_{\text{m},j}(\boldsymbol{k})$ denotes the $n_{th}$ member of perturbed surface elevation in Fourier space, and $\omega(\boldsymbol{k})$ is the angular frequency corresponding to the vector wavenumber $\boldsymbol{k}$.

Although the error statistics of the measurements can be fully determined by \eqref{eq:noise1}~and~\eqref{eq:lpsi}, it is a common practice in EnKF to compute the error covariance matrix directly from the ensemble \eqref{eq:etaeng} (in order to match the same procedure which has to be used for the forecast ensemble). For this purpose, we define an operator $\mathfrak{C}$ applied on ensemble $\boldsymbol{S}$ (such as $\boldsymbol{S}_{\text{m},j}$) such that
\begin{equation}
    \mathfrak{C}(\boldsymbol{S})=\boldsymbol{S}'(\boldsymbol{S}')^{\text{T}},
\end{equation}
where 
\begin{equation}
\boldsymbol{S}'=\frac{1}{\sqrt{N-1}}[s^{(1)}-\bar{s},~s^{(2)}-\bar{s},~\dots~s^{(n)}-\bar{s}~\dots ~s^{(N-1)}-\bar{s},~s^{(N)}-\bar{s}],
\label{eq:mean1}
\end{equation}
\begin{equation}
\bar{s}=\frac{1}{N}\sum_{n=1}^{N} s^{\text(n)}.
\end{equation}
Therefore, applying $\mathfrak{C}$ on $\boldsymbol{S}_{\text{m},j}$,
\begin{equation}
    \boldsymbol{R}_{s,j}=\mathfrak{C}(\boldsymbol{S}_{\text{m},j})
\label{eq:coper}
\end{equation}
gives the error covariance matrix of the measurements. 

\subsection{Nonlinear Wave Model by HOS}
\label{sec:hos}
Given the initial condition $s_{\text{m},j}^{(n)}$ for each ensemble member $n$, the evolution of $s^{(n)}({\boldsymbol{x}},t)$ is solved by integrating the surface wave equations in Zakharov form~\citep{zakharov1968stability} formulated as
\begin{equation}
\frac{\partial\eta(\boldsymbol{x},t)}{\partial t} + \frac{\partial\psi(\boldsymbol{x},t)}{\partial \boldsymbol{x}} \cdot \frac{\partial\eta(\boldsymbol{x},t)}{\partial \boldsymbol{x}} -\left[1+\frac{\partial\eta(\boldsymbol{x},t)}{\partial \boldsymbol{x}}\cdot \frac{\partial\eta(\boldsymbol{x},t)}{\partial \boldsymbol{x}}\right]\phi_z(\boldsymbol{x},t)=0,
\label{eq:bc1}
\end{equation}
\begin{equation}
\frac{\partial\psi(\boldsymbol{x},t)}{\partial t} + \frac{1}{2}\frac{\partial\psi(\boldsymbol{x},t)}{\partial \boldsymbol{x}}\cdot\frac{\partial\psi(\boldsymbol{x},t)}{\partial \boldsymbol{x}}+\eta(\boldsymbol{x},t)-\frac{1}{2}\left[1+\frac{\partial\eta(\boldsymbol{x},t)}{\partial \boldsymbol{x}}\cdot \frac{\partial\eta(\boldsymbol{x},t)}{\partial \boldsymbol{x}}\right]\phi_z(\boldsymbol{x},t)^2=0,
\label{eq:bc2}
\end{equation}
where $\phi_z(\boldsymbol{x},t)\equiv \partial \phi/\partial z|_{z=\eta}(\boldsymbol{x},t)$ is the surface vertical velocity with $\phi(\boldsymbol{x},z,t)$ being the velocity potential of the flow field. In \eqref{eq:bc1} and \eqref{eq:bc2}, we have assumed, for simplicity, that the time and mass units are chosen so that the gravitational acceleration and fluid density are unity \cite[e.g.][]{dommermuth1987high}.

The key procedure in HOS is to solve for $\phi_z(\boldsymbol{x},t)$ given $\psi(\boldsymbol{x},t)$ and $\eta(\boldsymbol{x},t)$, formulated as a boundary value problem for $\phi(\boldsymbol{x},z,t)$. This is achieved through a pseudo-spectral method in combination with a mode-coupling approach, with details included in multiple papers such as \cite{dommermuth1987high,pan2018high}.

\subsection{Data Assimilation Scheme by EnKF}
\label{sec:enkf}
Equations \eqref{eq:bc1} and \eqref{eq:bc2} are integrated in time for each ensemble member to provide the ensemble of forecasts at $t=t_j$ (for $j\geq1$):
\begin{equation}
\boldsymbol{S}_{\text{f},j}=[s_{\text{f},j}^{(1)},s_{\text{f},j}^{(2)},\cdots s_{\text{f},j}^{(n)},\cdots s_{\text{f},j}^{(N-1)},s_{\text{f},j}^{(N)}]\in \mathbb{R}^{L \times N},
\label{eq:etaen}
\end{equation}
where $L$ is the number of elements in the forecast state vector and $s_{\text{f},j}^{(n)}(\boldsymbol{x})\equiv s_{\text{f}}^{(n)}(\boldsymbol{x},t_j)$ is the $n_{th}$ member of the ensembles of model forecast results. The error covariance matrix of the model forecast can be computed by applying the operator $\mathfrak{C}$ on $\boldsymbol{S}_{\text{f},j}$:
\begin{equation}
\boldsymbol{Q}_{s,j}=\mathfrak{C}(\boldsymbol{S}_{\text{f},j}).
\label{eq:covetaf}
\end{equation}

An analysis step is then performed, which combines the ensembles of model forecasts and perturbed measurements to produce the optimal analysis results~\citep{carrassi2018data}:
\begin{equation}
\mathop{\boldsymbol{S}_{\text{a},j}}_{L \times N}=\mathop{\boldsymbol{S}_{\text{f},j}}_{L \times N}+\mathop{\boldsymbol{K}_{s,j}}_{L\times d}[\mathop{\boldsymbol{S}_{\text{m},j}}_{d\times N}-\mathop{\boldsymbol{G}_j}_{d\times L} \mathop{\boldsymbol{S}_{\text{f},j}}_{L\times N}],
\label{eq:ana1}
\end{equation}
where 
\begin{equation}
  {\boldsymbol{K}_{s,j}}={\boldsymbol{Q}_{s,j}}{\boldsymbol{G}_j^\text{T}}\big[\boldsymbol{G}_j\boldsymbol{Q}_{s,j}\boldsymbol{G}_j^\text{T}+\boldsymbol{R}_{s,j}]^{-1} 
\label{eq:k1}
\end{equation}
is the optimal Kalman gain matrix of the state (for $s$=$\eta$ or $s$=$\psi$). $\boldsymbol{G}_j$ is a linear operator, which maps a state vector from the model space to the measurement space: $\mathbb{R}^L\to\mathbb{R}^{d}$. In the present study, $\boldsymbol{G}_j$ is constructed by considering a linear interpolation (or Fourier interpolation~\citep{grafakos2008classical}) from the space of model forecast, i.e., $s_{\text{f},j}^{(n)}$, to the space of measurements, i.e., $s_{\text{m},j}^{(n)}$. 

While we have now completed the formal introduction of the EnKF-HOS algorithm (and all steps associated with figure \ref{fig:enkfhos}), additional procedures are needed to improve the robustness of EnKF and address the possible mismatch between the predictable zone and measurement region. These will be discussed respectively in \S2.6 and \S2.7, with the former leading to a (heuristic but effective) correction of $\boldsymbol{S}_{\text{f},j}$ and $\boldsymbol{Q}_{s,j}$ before~\eqref{eq:ana1} and \eqref{eq:k1} are applied, and the latter a modification of \eqref{eq:ana1} when the mismatch occurs.

\subsection{Adaptive Inflation and Localization}
With $N\rightarrow \infty$ and exact representation of physics by \eqref{eq:bc1} and \eqref{eq:bc2}, it is expected that \eqref{eq:etaen} and \eqref{eq:covetaf} capture the accurate statistics of the model states and equation~\eqref{eq:ana1} provides the true optimal analysis. However, due to the finite ensemble size and the underrepresented physics in \eqref{eq:bc1} and \eqref{eq:bc2}, errors associated with statistics computed by \eqref{eq:covetaf} may lead to sub-optimal analysis and even (classical) filter divergence \citep{evensen2003ensemble,evensen2009data,carrassi2018data}. These errors have been investigated by numerous previous studies \citep{houtekamer2005atmospheric,lorenc2003potential,hansen2002accounting,hamill2005accounting,evensen2003ensemble,evensen2009data, carrassi2018data}, which are characterized by (i) underestimates of the ensemble variance in $\boldsymbol{Q}_{s,j}$ and (ii) spurious correlations in $\boldsymbol{Q}_{s,j}$ over long spatial distances. To remedy this situation, adaptive inflation and localization (respectively for error (i) and (ii)) are usually applied as common practices in EnKF to correct $\boldsymbol{S}_{\text{f},j}$ and $\boldsymbol{Q}_{s,j}$ before they are used in \eqref{eq:ana1} and~\eqref{eq:k1}.

In this work, we apply the adaptive inflation algorithm \citep{anderson1999monte,anderson2007adaptive} in our EnKF-HOS framework. Specifically, each ensemble member in \eqref{eq:etaen} is linearly inflated before the subsequent computation, i.e.,
\begin{equation}
    s^{(n),\text{inf}}_{\text{f},j}=\sqrt{\lambda_j}(s^{(n)}_{\text{f},j}-\bar{s}_{\text{f},j})+\bar{s}_{\text{f},j},~n=1,2\cdots~N,
    \label{eq:inflationeta}
\end{equation}
where $\lambda_j\geq 1$ is referred to as the covariance inflation factor. The purpose of \eqref{eq:inflationeta} is to amplify the underestimated ensemble variance in $\boldsymbol{Q}_{s,j}$, especially when $\bar{s}_{\text{f},j}$ is far from $s_{\text{m},j}$, therefore to avoid the ignorance of $\boldsymbol{S}_{\text{m},j}$ in \eqref{eq:ana1} (i.e., filter divergence) due to the overconfidence in the forecast. The appropriate value of $\lambda_j$ can be determined at each $t=t_j$ through the adaptive inflation algorithm~\citep{anderson2007adaptive}, which considers $\lambda_j$ as an additional state variable maximizing a posterior distribution $p(\lambda_j|\eta_{\text{m},j})$. The detailed algorithm is presented in Appendix~\ref{appA}.

After obtaining the inflated $\boldsymbol{Q}_{s,j}$, a localization scheme is applied, which removes the spurious correlation by performing the Schur product (i.e., element-wise matrix product) between $\boldsymbol{Q}_{s,j}$ and a local-correlation function $\bm{\mu}$~\citep{carrassi2018data}
\begin{equation}
    \boldsymbol{Q}^{loc}_{s,j}=\bm{\mu}\circ\boldsymbol{Q}_{s,j}.
    \label{eq:loc}
\end{equation}
with $\bm{\mu}$ defined as the Gaspari–Cohn (GC) function (see Appendix~\ref{appB} for details).

\subsection{Interplay between predictable zone and measurement region}
\begin{figure}
  \centerline{\includegraphics[scale =0.7]{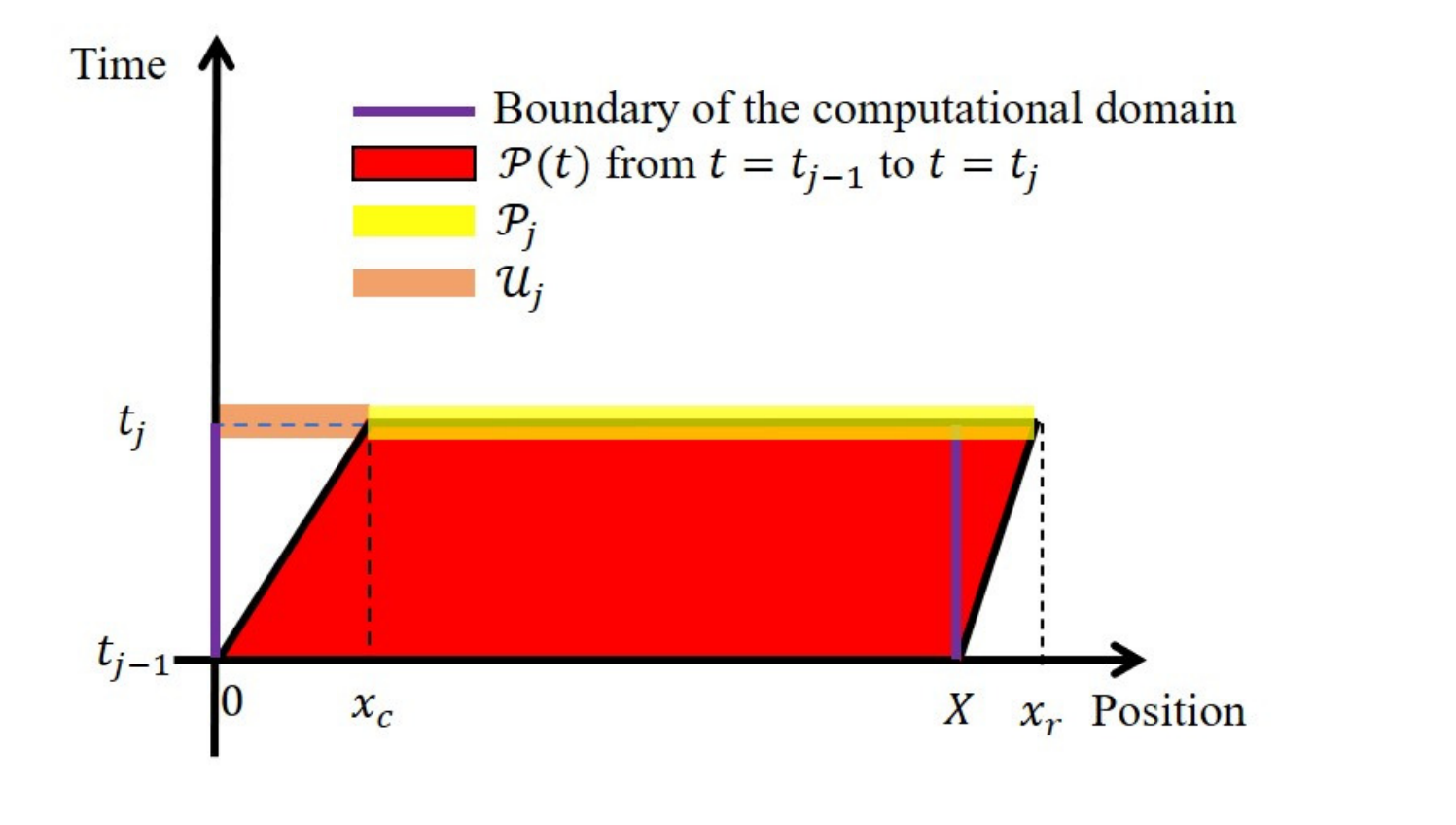}}
  \caption{The spatial-temporal predictable zone (red) for a case of uni-directional waves traveling to the right, with initial data in $[0,X]$ at $t=t_{j-1}$. The boundary of the computational domain is indicated by the purple vertical lines. The left and right boundary moves with speed $c_{\text{g}}^{\text{max}}$ and $c_{\text{g}}^{\text{min}}$ respectively. At $t=t_j$, the spatial predictable zone $\mathcal{P}_j$ (yellow) is located in $[x_c,x_r]=[c_{\text{g}}^{\text{max}}(t_j-t_{j-1}), X+c_{\text{g}}^{\text{min}}(t_j-t_{j-1})]$ (or in $[x_c, X]$ within the computational domain), and the unpredictable zone $\mathcal{U}_j$ (orange) is located in $[0,x_c]$.}
\label{fig:pz}
\end{figure}

The predictable zone is a spatio-temporal zone where the wave field is computationally tractable given an observation of the field in a limited space at a specific time instant~\citep{naaijen2014limits,kollisch2018nonlinear,qi2018predictable}. Depending on the wave travelling direction, the boundary of the spatial predictable zone moves in time with speed $c_{\text{g}}^{\text{max}}$ or $c_{\text{g}}^{\text{min}}$, which are the maximum and minimum group speeds within all wave modes of interest (see figure~\ref{fig:pz} for an example of predictable zone $\mathcal{P}(t)$ and unpredictable zone $\mathcal{U}(t)$ for a uni-directional wave field starting with initial data in $[0,X]$). 

In practice, for a forecast from $t_{j-1}$ to $t_j$, the predictable zone $\mathcal{P}(t)$ at $t=t_j$ only constitutes a sub-region of the computational domain (see caption of figure~\ref{fig:pz}), and there is no guarantee that the measurement region $\mathcal{M}_j$ overlaps with the predictable zone $\mathcal{P}_j=\mathcal{P}(t_j)$. This requires a special treatment of the region where the measurements are available but the forecast is untrustworthy, i.e., $\boldsymbol{x}\in (\mathcal{U}_j\cap\mathcal{M}_j)$, where $\mathcal{U}_j=\mathcal{U}(t_j)$. To address this issue, we develop a modified analysis equation which replaces \eqref{eq:ana1} when considering the interplay among $\mathcal{P}_j$, $\mathcal{U}_j$ and $\mathcal{M}_j$.   

We consider our computational region as a subset of $\mathcal{P}_j\cup\mathcal{M}_j$, so that the analysis results at all $\boldsymbol{x}$ can be determined from the prediction and/or measurements. We further partition the forecast and analysis state vectors $s_{\text{f},j}^{(n)}$ and $s_{\text{a},j}^{(n)}$ (in the computational domain), as well as the measurement state vector $s_{\text{m},j}^{(n)}$ (in $\mathcal{M}_j$), according to 
\begin{equation}
    \boldsymbol{S}_{*,j}\in\mathbb{R}^{L\times N} =
    \begin{bmatrix}

    \boldsymbol{S}^{\mathcal{P}}_{*,j}\in\mathbb{R}^{L^{\mathcal{P}}\times N}\\
    \boldsymbol{S}^{\mathcal{U}}_{*,j}\in\mathbb{R}^{L^{\mathcal{U}}\times N}
    \end{bmatrix}
    ,
    \boldsymbol{S}_{m,j}\in\mathbb{R}^{d\times N} =
    \begin{bmatrix}
    \boldsymbol{S}^{\mathcal{P}}_{m,j}\in\mathbb{R}^{d^{\mathcal{P}}\times N}\\
    \boldsymbol{S}^{\mathcal{U}}_{m,j}\in\mathbb{R}^{d^{\mathcal{U}}\times N}
    \end{bmatrix}
\label{eq:modenkf1}
\end{equation}
where $*=\text{f},\text{a}$. The variables with superscript $\mathcal{U}$ ($\mathcal{P}$) represent the part of state vectors for which $\boldsymbol{x}\in \mathcal{U}_j$ ($\boldsymbol{x}\in \mathcal{P}_j$), with associated number of elements $L^{\mathcal{U}}$ and $d^{\mathcal{U}}$ ($L^{\mathcal{P}}$ and $d^{\mathcal{P}}$) for $s_{*,j}^{(n)}$ and $s_{\text{m},j}^{(n)}$. The modified analysis equation is formulated as

\begin{subequations}
\begin{align}
\mathop{\boldsymbol{S}^{\mathcal{P}}_{\text{a},j}}_{L^\mathcal{P}\times N}&=\mathop{\boldsymbol{S}^{\mathcal{P}}_{\text{f},j}}_{L^\mathcal{P}\times N}+\mathop{\boldsymbol{K}^{\mathcal{P}}_{s,j}}_{L^\mathcal{P}\times d^\mathcal{P}}[\mathop{\boldsymbol{S}^{\mathcal{P}}_{\text{m},j}}_{d^\mathcal{P}\times N}-\mathop{\boldsymbol{G}^{\mathcal{P}}_j}_{d^\mathcal{P}\times L^\mathcal{P}} \mathop{\boldsymbol{S}^{\mathcal{P}}_{\text{f},j}}_{L^\mathcal{P}\times N}],
\label{eq:anap}\\
\mathop{\boldsymbol{S}^{\mathcal{U}}_{\text{a},j}}_{L^\mathcal{U}\times N}&=\mathop{\mathcal{H}_j}_{L^{\mathcal{U}}\times d}\mathop{\boldsymbol{S}_{\text{m},j}}_{d\times N},
\label{eq:anau}
\end{align}
\label{eq:ananew}
\end{subequations}
where $\boldsymbol{K}^{\mathcal{P}}_{s,j}= \boldsymbol{K}_{s,j}(1:L^{\mathcal{P}},1:d^{\mathcal{P}})$ and $\boldsymbol{G}^{\mathcal{P}}_{s,j}= \boldsymbol{G}_{s,j}(1:d^{\mathcal{P}},1:L^{\mathcal{P}})$ are sub-matrices of $\boldsymbol{K}_{s,j}$ and $\boldsymbol{G}_{s,j}$ associated with $\boldsymbol{x}\in \mathcal{P}_j$ in both measurement and forecast (analysis) spaces, $\mathcal{H}_j$ is a linear operator which maps a state vector from measurement space to unpredictable zone in the analysis space: $\mathbb{R}^d\to\mathbb{R}^{L^\mathcal{U}}$ (based on linear/Fourier interpolation). 

By implementing~\eqref{eq:anap}, $\boldsymbol{S}^{\mathcal{U}}_{\text{f},j}$ is discarded in the analysis to compute $\boldsymbol{S}^{\mathcal{P}}_{\text{a},j}$; and by \eqref{eq:anau}, $\boldsymbol{S}^{\mathcal{U}}_{\text{a},j}$ is determined only from the measurements $\boldsymbol{S}_{\text{m},j}$ without involving $\boldsymbol{S}^{\mathcal{U}}_{\text{f},j}$. Therefore, the modified EnKF analysis equation provides the minimum analysis error when considering the interplay among $\mathcal{P}_j$, $\mathcal{U}_j$ and $\mathcal{M}_j$.

\subsection{Pseudo-code and computational cost}
Finally, we provide a pseudo-code for the complete EnKF-HOS coupled algorithm in Algorithm~\ref{al:asimilation}. For ensemble forecast by HOS, the algorithm takes $\mathcal{O}(NLlogL)$ operations for each time step $\Delta t$. For the analysis step at $t= t_j$, the algorithm has a computational complexity of $\mathcal{O}(dLN)$ for \eqref{eq:ana1} and $\mathcal{O}(dL^2)+\mathcal{O}(d^3)+\mathcal{O}(d^2L)$ for \eqref{eq:k1} (if Gaussian elimination is used for the inverse). Therefore, the average computational complexity for one time step is $\mathcal{O}(NLlogL)+\mathcal{O}(dL^2)/(\tau/\Delta t)$ (for $L>\sim N$ and $L>\sim d$), with $\tau$ the DA interval.

\begin{algorithm*}[htpb!]
\centering
\caption{Algorithm for EnKF-HOS method}

\label{al:asimilation}
\begin{algorithmic}[1]
\State {\bf{Input}}: ${\eta_{\text{m},0}}$, ${\psi_{\text{m},0}}$ (initial measurements), $t_{\text{max}}$ (final computational time), $N$ (ensemble size), $\mathcal{T}=\{t_1, t_2, t_3\cdots\}$ (time instants of data assimilation), $\bar{\lambda}_0$ and $\sigma^2_{0}$ (initial guess of the inflation factor, see Appendix A)
\State {\bf{Begin}}
\State initialize:\\
\hspace{1.2cm}       $t=t_0, j=0$\\
\hspace{1.2cm}       Generate $\boldsymbol{S}_{\text{m},0}$ with \eqref{eq:etaeng} $\sim$ \eqref{eq:lpsi}.
\State time loop:
\State \hspace{1.2cm} {\bf{while}} $t \leq t_{\text{max}} $ {\bf{do}}\
\State \hspace{1.6cm} $j=j+1$.\
\State \hspace{1.6cm} Solve \eqref{eq:bc1} and \eqref{eq:bc2} until $t=t_j$ to obtain $\boldsymbol{S}_{\text{f},j}$.
\State \hspace{1.6cm} {\bf{read}} $\eta_{\text{m},j}$ (measurements)
\State \hspace{1.6cm} Generate $\boldsymbol{S}_{\text{m},j}$ with \eqref{eq:etaeng} $\sim$ \eqref{eq:lpsi}.\
\State \hspace{1.6cm} Calculate $\boldsymbol{R}_{s,j}$ with~\eqref{eq:coper}.
\State \hspace{1.6cm} Perform adaptive inflation with \eqref{eq:inflationeta} and calculate $\boldsymbol{Q}_{s,j}$ with~\eqref{eq:covetaf}.
\State \hspace{1.6cm} Perform covariance localization on $\boldsymbol{Q}_{s,j}$ with~\eqref{eq:loc}. 
\State \hspace{1.6cm} Calculate $\boldsymbol{S}_{\text{a},j}$ with \eqref{eq:ana1} (or \eqref{eq:ananew} when considering $\mathcal{P}_j$, $\mathcal{U}_j$, and $\mathcal{M}_j$).
\State \hspace{1.6cm} {\bf{Output}} $\bar{s}_{\text{a},j}$ (ensemble mean of $\boldsymbol{S}_{\text{a},j}$).
\State \hspace{1.6cm}  $\boldsymbol{S}_{\text{f},j}=\boldsymbol{S}_{\text{a},j}$.
\State \hspace{1.2cm} {\bf{end}}
\State {\bf{end}}
\end{algorithmic}
\end{algorithm*}

\section{Numerical results}
\label{sec:numres}
To test the performance of the EnKF-HOS algorithm, we apply it on a series of cases with both synthetic and real ocean wave fields. For the former, we use a reference HOS simulation to generate the true wave field, on top of which we superpose random errors to generate the synthetic noisy measurements. For the latter, we use real data collected by a ship-borne Doppler coherent marine radar \citep{lyzenga2015real,nwogu2010surface}. The adaptive inflation and localization algorithms are only applied for the latter case, where the under-represented physics in \eqref{eq:bc1} and \eqref{eq:bc2} significantly influences the model statistics. The perturbed measurement ensemble \eqref{eq:etaeng} are generated with parameters $c=0.0025 \sigma^2_{\eta}$ (where $\sigma_{\eta}$ is the standard deviation of the surface elevation field) and $a=\lambda_0/8$ (where $\lambda_0$ is the fundamental wavelength in the computational domain) in \eqref{eq:noise1} in all cases unless otherwise specified. We remark that these choices may not reflect the true error statistics of the radar measurement, which unfortunately has not been characterized yet. The results from EnKF-HOS simulations are compared with HOS-only simulations (both taking noisy measurements as initial conditions) to demonstrate the advantage of the new EnKF-HOS method.

\subsection{Synthetic cases}
We consider the synthetic cases where the true solution of a wave field ($\eta^{\text{true}}(\boldsymbol{x},t)$ and~$\psi^{\text{true}}(\boldsymbol{x},t)$) is generated by a single reference HOS simulation starting from the (exact) initial condition. The (noisy) measurements of surface elevation are generated by superposing random error on the true solution:
\begin{equation}
    \eta_{\text{m},j}(\boldsymbol{x})=\eta^\text{true}(\boldsymbol{x},t_j)+v(\boldsymbol{x}),~j=0,1,2\cdots,
    \label{eq:mea}
\end{equation}
where $v(\boldsymbol{x})$ is a random field, which represents the measurement error and shares the same distribution as $w^{(n)}$ (see~\eqref{eq:noise1}). For simplicity in generating initial model ensemble, we use $\eta_{\text{m},0}\in\mathbb{R}^L$ in \eqref{eq:mea} (i.e., $d_0=L$ in \eqref{eq:etaeng}), and $\eta_{\text{m},j}\in\mathbb{R}^d$ for $j\geq 1$ with $d$ specified in each case below. Similar to $\psi^{(n)}_{\text{m},j}(\boldsymbol{x})$, $\psi_{\text{m},j}(\boldsymbol{x})$ is generated based on the linear wave theory,
\begin{equation}
\psi_{\text{m},j}(\boldsymbol{x})=\int\frac{i\omega(\boldsymbol{k})}{|\boldsymbol{k}|}\tilde{\eta}_{\text{m},j}(\boldsymbol{k})e^{i\boldsymbol{k}\cdot\boldsymbol{x}}d\boldsymbol{k},
\label{eq:e2}
\end{equation}
where $\tilde{\eta}_{\text{m},j}(\boldsymbol{k})$ denotes the measurement of surface elevation in Fourier space.



Depending on how the true solution is generated, we further classify the synthetic cases into idealistic and realistic cases. In the idealistic case, the true solution is taken from an HOS simulation with periodic boundary conditions, so that the entire computational domain is predictable. In the realistic case, we consider the true solution as a patch in the boundless ocean (practically taken from a patch in a much larger domain where the HOS simulation is conducted), and the interplay between $\mathcal{M}_j$ and $\mathcal{P}_j$ discussed in \S 2.7 is critical. Correspondingly, we apply the modified EnKF analysis equation \eqref{eq:ananew} only in the realistic cases. 

To quantify the performance of EnKF-HOS and HOS-only methods, we define an error metric 
\begin{equation}
\epsilon(t;\mathcal{A})=\frac{\int_\mathcal{A}\mid\eta^\text{true}(\boldsymbol{x},t)-\eta^{\text{sim}}(\boldsymbol{x},t)\mid^2 d\mathcal{A}}{2\sigma_{\eta}^2 \mathcal{A}},
\label{eq:epsilon}
\end{equation}
where $\mathcal{A}$ is a region of interest based on which the spatial average is performed (here we use $\mathcal{A}$ to represent both the region and its area), $\eta^{\text{sim}}(\boldsymbol{x},t)$ represents the simulation results obtained from EnKF-HOS (the ensemble average in this case) or the HOS-only method, and $\sigma_\eta$ is the standard deviation of, say, $\eta^{\text{true}}$ in $\mathcal{A}$. It can be shown that the definition \eqref{eq:epsilon} yields $\epsilon(t;\mathcal{A})=1-\rho_{\mathcal{A}}(\eta^\text{true},\eta^\text{sim})$, with
\begin{equation}
    \rho_{\mathcal{A}}(\eta^\text{a},\eta^\text{b})=\frac{\int_\mathcal{A}\eta^\text{a}(\boldsymbol{x},t)\eta^\text{b}(\boldsymbol{x},t) d\mathcal{A}}{\sigma_{\eta}^2 \mathcal{A}},
    \label{eq:corr}
\end{equation}
being the correlation coefficient between $\eta^\text{a}$ and $\eta^\text{b}$ (in this case $\eta^{\text{true}}$ and $\eta^\text{sim}$). Therefore, $\epsilon(t;\mathcal{A})=1$ corresponds to the case that all phase information is lost in the simulation.

In the following, we show results for idealistic and realistic cases of synthetic irregular wave fields. Preliminary results validating the EnKF-HOS method for Stokes waves in the idealistic setting can be found in \cite{wangdata}.

\subsubsection{Results for idealistic cases}
\label{sec:ideal}
We consider idealistic cases with both two-dimensional (2D, with one horizontal direction $x$) and three-dimensional (3D, with two horizontal directions $\boldsymbol{x}=(x,y)$) wave fields. The true solutions for both situations are obtained from reference simulations starting from initial conditions prescribed by a realization of the JONSWAP spectrum $S(\omega)$, with a spreading function $D(\theta)$ for the 3D case (where $\omega$ and $\theta$ are the angular frequency and angle with respect to the positive $x$ direction).

\begin{figure}
     \centering
     \begin{subfigure}[b]{1\textwidth}
         \centering
         \includegraphics[trim=2cm 0cm 4cm 0cm, clip,scale=0.4]{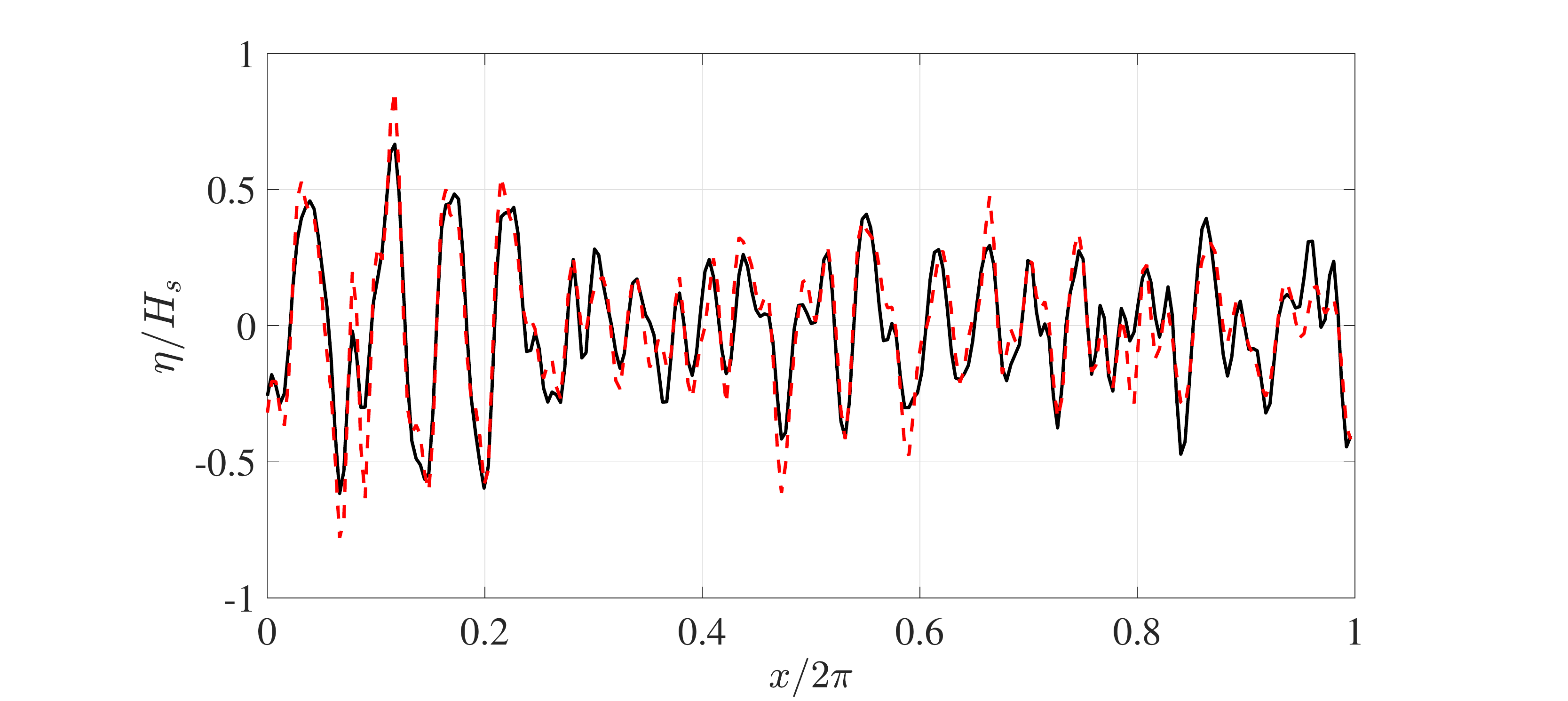}
         \caption{}
         \label{fig:ic_eta}
     \end{subfigure}
     \begin{subfigure}[b]{1\textwidth}
         \centering
         \includegraphics[trim=2cm 0cm 4cm 0cm, clip,scale=0.4]{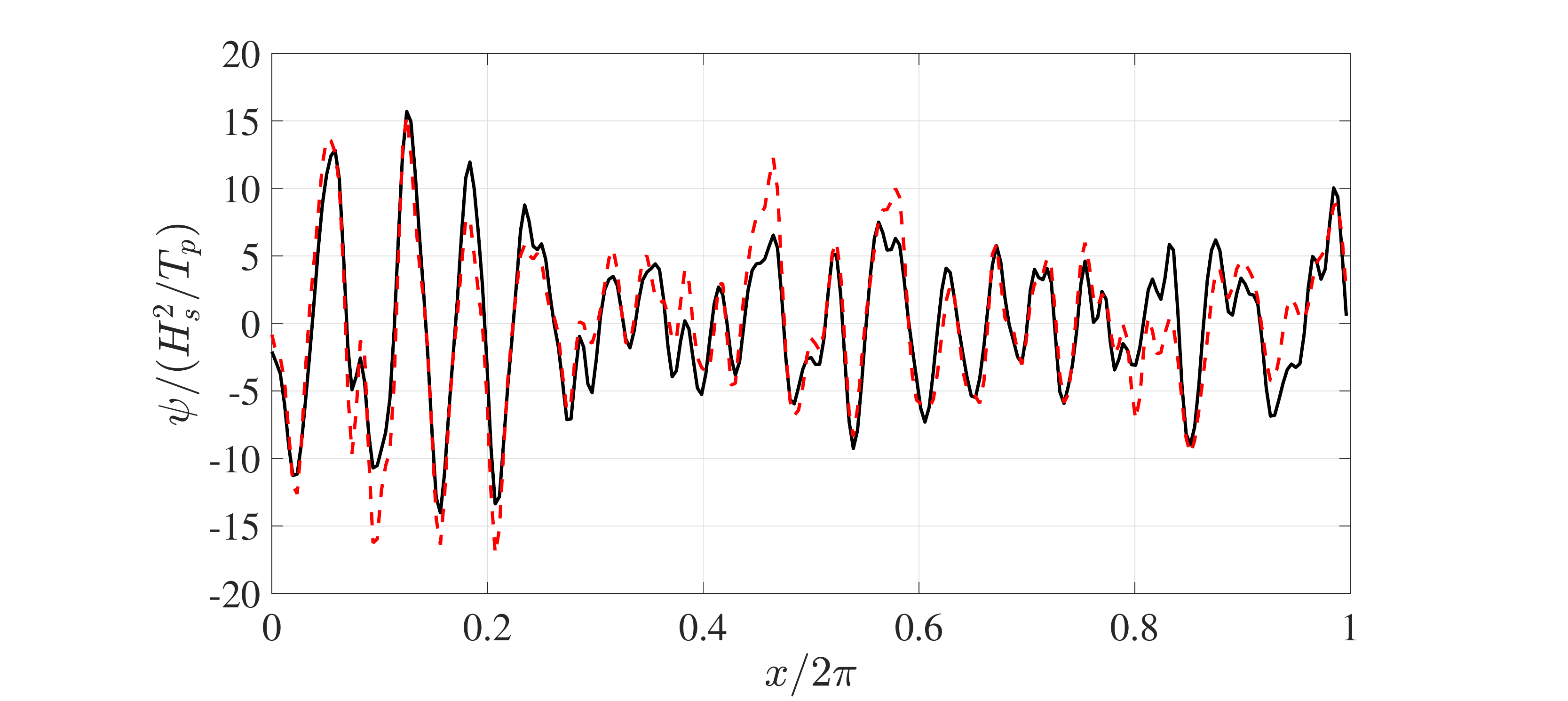}
         \caption{}
         \label{fig:ic_psi}
     \end{subfigure}
        \caption{Plots of (a) $\eta^{\text{true}}(x,t_0)$ (\rule[0.5ex]{0.5cm}{0.25pt}) and $\eta_{\text{m},0}(x)$ ({\color{red}\dashL}); (b) $\psi^{\text{true}}(x,t_0)$ (\rule[0.5ex]{0.5cm}{0.25pt}) and $\psi_{\text{m},0}(x)$ ({\color{red}\dashL}).}
        \label{fig:ic2d}
\end{figure}

For the first case of the 2D wave field, we use an initial spectrum $S(\omega)$ with global steepness $k_pH_s/2=0.11$, peak wavenumber $k_p=16k_0$ with $k_0$ the fundamental wavenumber in the computational domain, and enhancement factor $\gamma=3.3$. $L=256$ grid points are used in spatial domain $[0,2\pi)$ in the (reference, EnKF-HOS and HOS-only) simulations. The noisy measurements $s_{\text{m},j}$ are generated through \eqref{eq:mea} and \eqref{eq:e2}, with a comparison between $s_{\text{m},0}(x)$ and $s^{\text{true}}(x,t_0)$ shown in figure \ref{fig:ic2d}. 

Both EnKF-HOS and HOS-only simulations start from initial measurements $s_{\text{m},0}(x)$. In the EnKF-HOS method, the ensemble size is set to be $N=100$, and measurements at $d=2$ locations of $x/(2\pi)=100/256$ and $170/256$ are assimilated into the model with a constant DA interval $\tau=t_j-t_{j-1}=T_p/16$, where $T_p=2\pi/\sqrt{k_p}$ from the dispersion relation.

\begin{figure}
  \centerline{\includegraphics[scale =0.4]{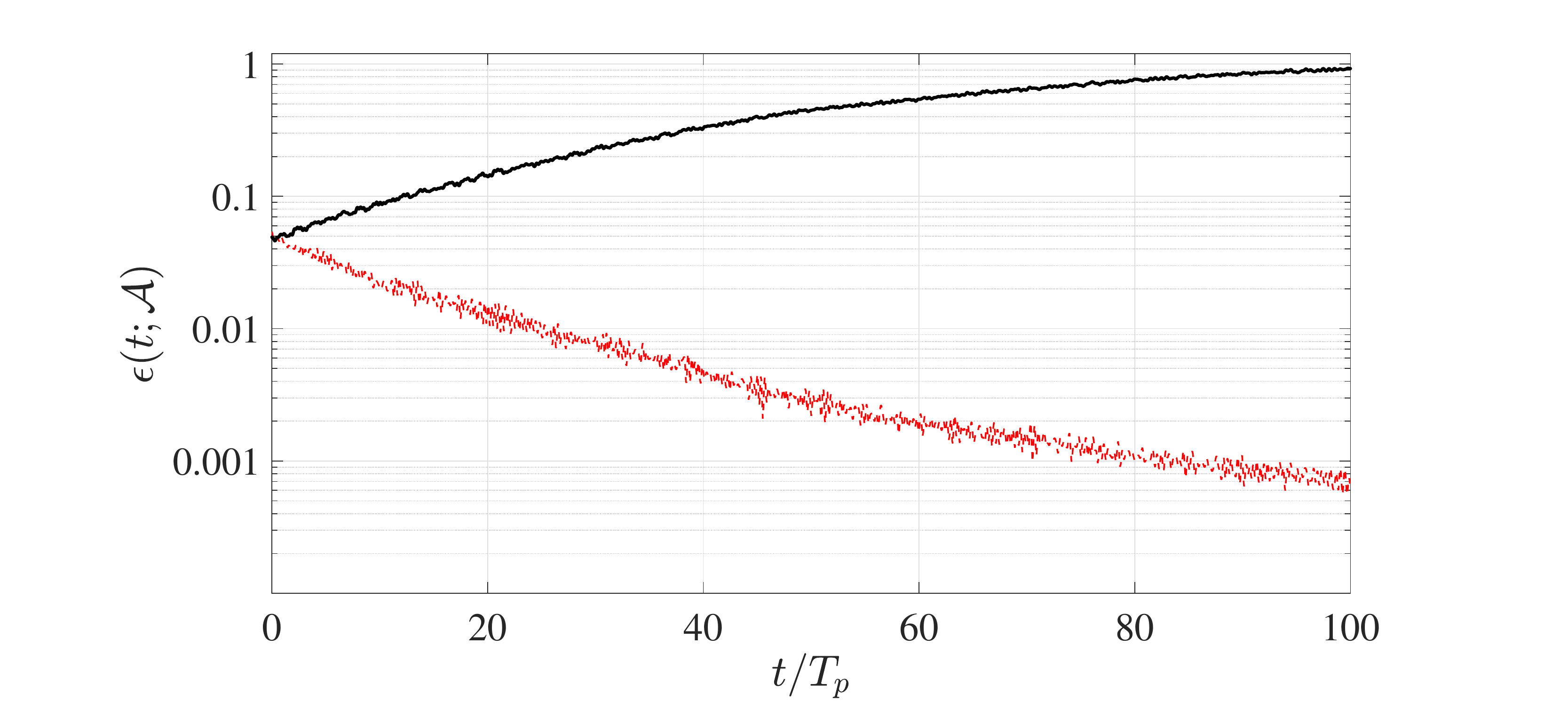}}
  \caption{Error $\epsilon(t;\mathcal{A})$ with $\mathcal{A}=[0,2\pi)$ from EnKF-HOS ({\color{red}\dashL}) and HOS-only (\rule[0.5ex]{0.5cm}{0.25pt}) methods, for the 2D idealistic case.}
\label{fig:compare_error_2d}
\end{figure}

\begin{figure}
     \centering
     \begin{subfigure}[b]{1\textwidth}
         \centering
         \includegraphics[trim=2cm 0cm 0cm 0cm, clip,scale=0.4]{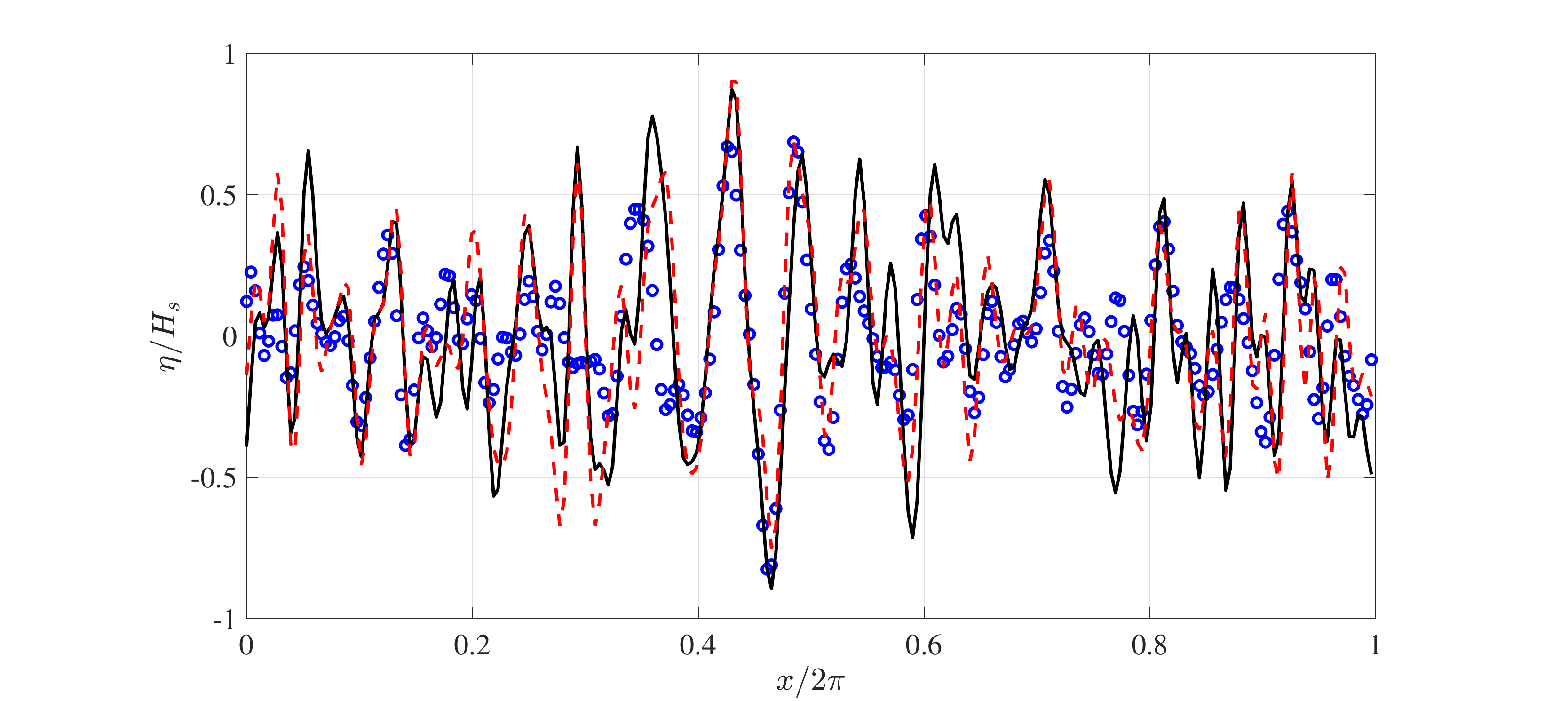}
         \caption{}
         \label{fig:snap1}
     \end{subfigure}
     \begin{subfigure}[b]{1\textwidth}
         \centering
         \includegraphics[trim=2cm 0cm 0cm 0cm, clip,scale=0.4]{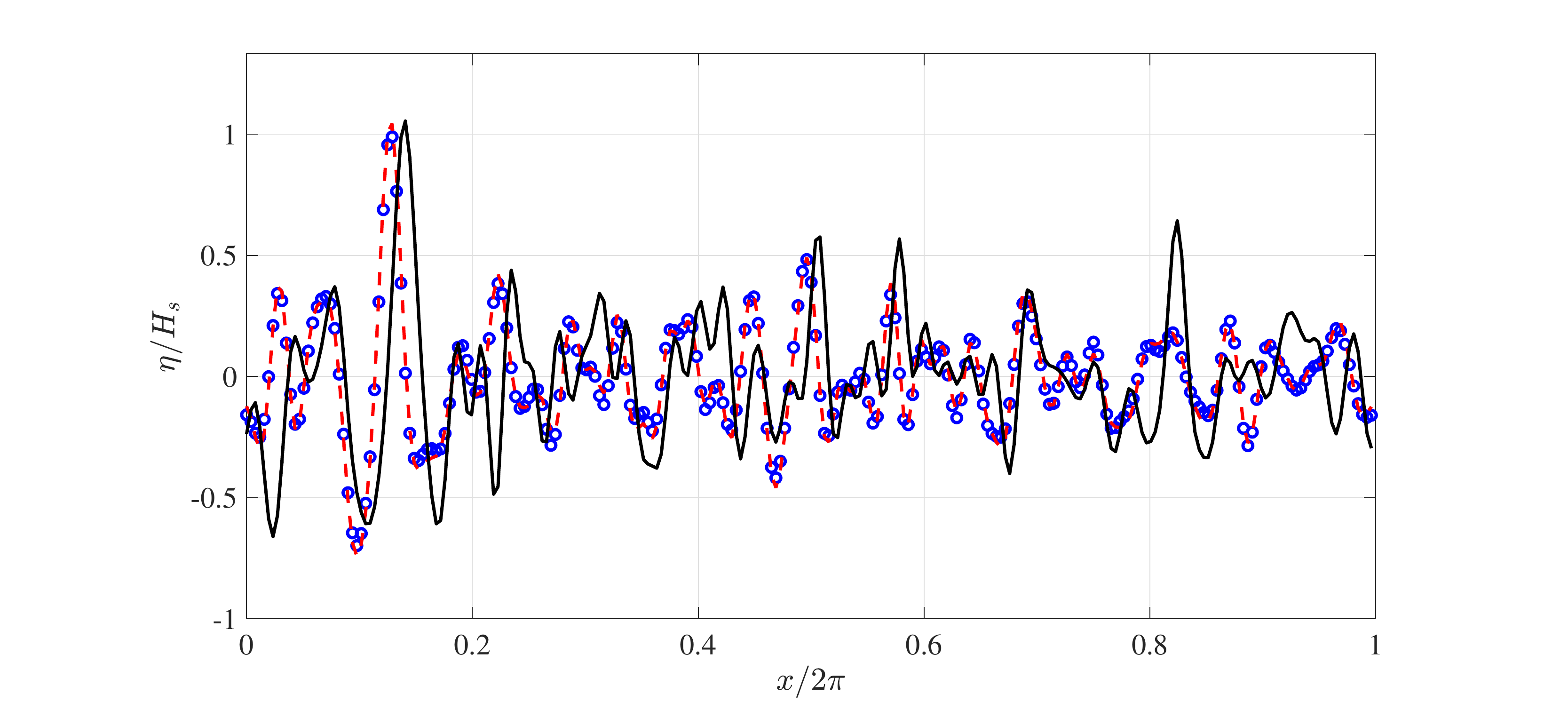}
         \caption{}
         \label{fig:snap1point5}
     \end{subfigure}
     
     \begin{subfigure}[b]{1\textwidth}
         \centering
         \includegraphics[trim=2cm 0cm 0cm 0cm, clip,scale=0.4]{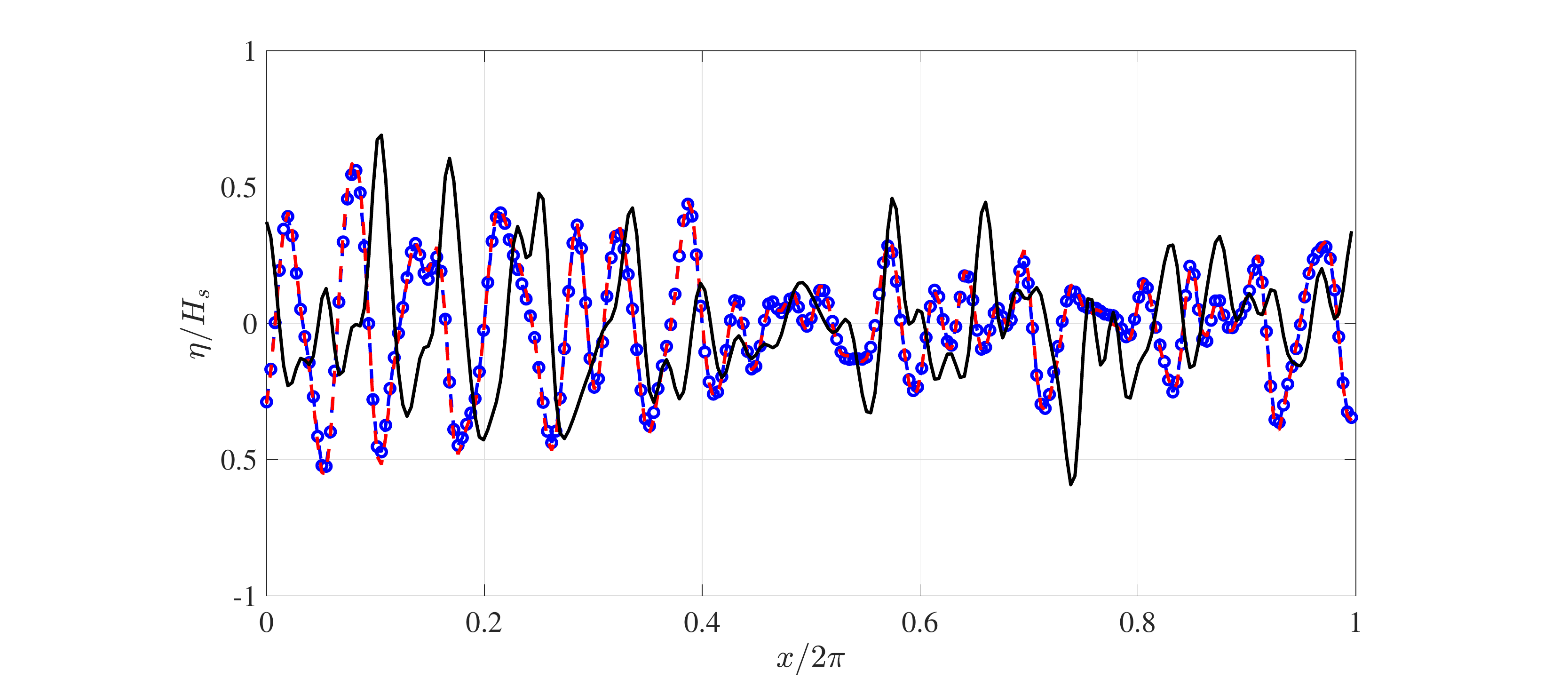}
         \caption{}
         \label{fig:snap2}
     \end{subfigure}
        \caption{Surface elevations $\eta^{\text{true}}(x)$ ({\color{blue}$\circ$}), $\eta^{\text{sim}}(x)$ with EnKF-HOS ({\color{red}\dashL}) and HOS-only (\rule[0.5ex]{0.5cm}{0.25pt}) methods, at (a) $t/T_p=5$, (b) $t/T_p=50$ and (c) $t/T_p=95$.}
        \label{fig:snaps}
\end{figure}

The error $\epsilon(t;\mathcal{A})$ with $\mathcal{A}=[0,2\pi)$ obtained from EnKF-HOS and HOS-only simulations are shown in figure~\ref{fig:compare_error_2d}. For HOS-only method, i.e., without DA, $\epsilon(t;\mathcal{A})$ increases in time from the initial value $\epsilon(0;\mathcal{A})\approx 0.05$ (see figure \ref{fig:ic2d}), and reaches $\mathcal{O}(1)$ at $t/T_p\approx 100$. In contrast, $\epsilon(t;\mathcal{A})$ from EnKF-HOS simulation keeps decreasing, and becomes several orders of magnitude smaller than that from HOS-only method (and two orders of magnitude smaller than the measurement error) at the end of the simulation. For visualization of the wave fields, figure~\ref{fig:snaps} shows snapshots of $\eta^{\text{true}}(x)$ and $\eta^{\text{sim}}(x)$ (with 
EnKF-HOS and HOS-only methods) at three time instants of $t/T_p=5,~45,~\text{and}~95$, which indicates the much better agreement with $\eta^{\text{true}}(x)$ when DA is applied. Notably, at and after $t/T_p=45$, the EnKF-HOS solution is not visually distinguishable from $\eta^\text{true}(x)$.

The influence of the parameter $c$ (reflecting the measurement error) on the results from both methods are summarized in Table~\ref{tab:e_steep}. We present the critical time instants $t^*$ when $\epsilon(t^*;\mathcal{A})$ reaches $\mathcal{O}(1)$ in HOS-only method, i.e. when the simulation completely loses the phase information. As expected, all cases lose phase information for sufficiently long time, and the critical time $t^*$ decreases with the increase of $c$. In contrast, for EnKF-HOS method, the error $\epsilon(t;\mathcal{A})$ decreases with time and reaches $\mathcal{O}(10^{-3})$ at $t=100T_p$ in all cases.

\begin{table}
  \begin{center}
\def~{\hphantom{0}}
  \begin{tabular}{lcc}
          &  $t^*$ & $\epsilon(100T_p;\mathcal{A})$ with EnKF-HOS 
          \\[3pt]
       $c=0.0004\sigma_{\eta}^2$   & $150T_p$ & ~~$1.65\times10^{-3}$~ \\
       $c=0.0025\sigma_{\eta}^2$   & $100T_p$ & ~~$6.21\times10^{-3}$~\\
       $c=0.0100\sigma_{\eta}^2$  & $70T_p$ & ~~$7.28\times10^{-3}$~ \\
       $c=0.0400\sigma_{\eta}^2$  & $40T_p$ & ~~$9.02\times10^{-3}$~ \\
  \end{tabular}
  \caption{Values of $t^*$ in HOS-only method and $\epsilon(100T_p;\mathcal{A})$ in EnKF-HOS method for different values of $c$.}
  \label{tab:e_steep}
  \end{center}
\end{table}

\begin{figure}
  \centerline{\includegraphics[scale =0.4]{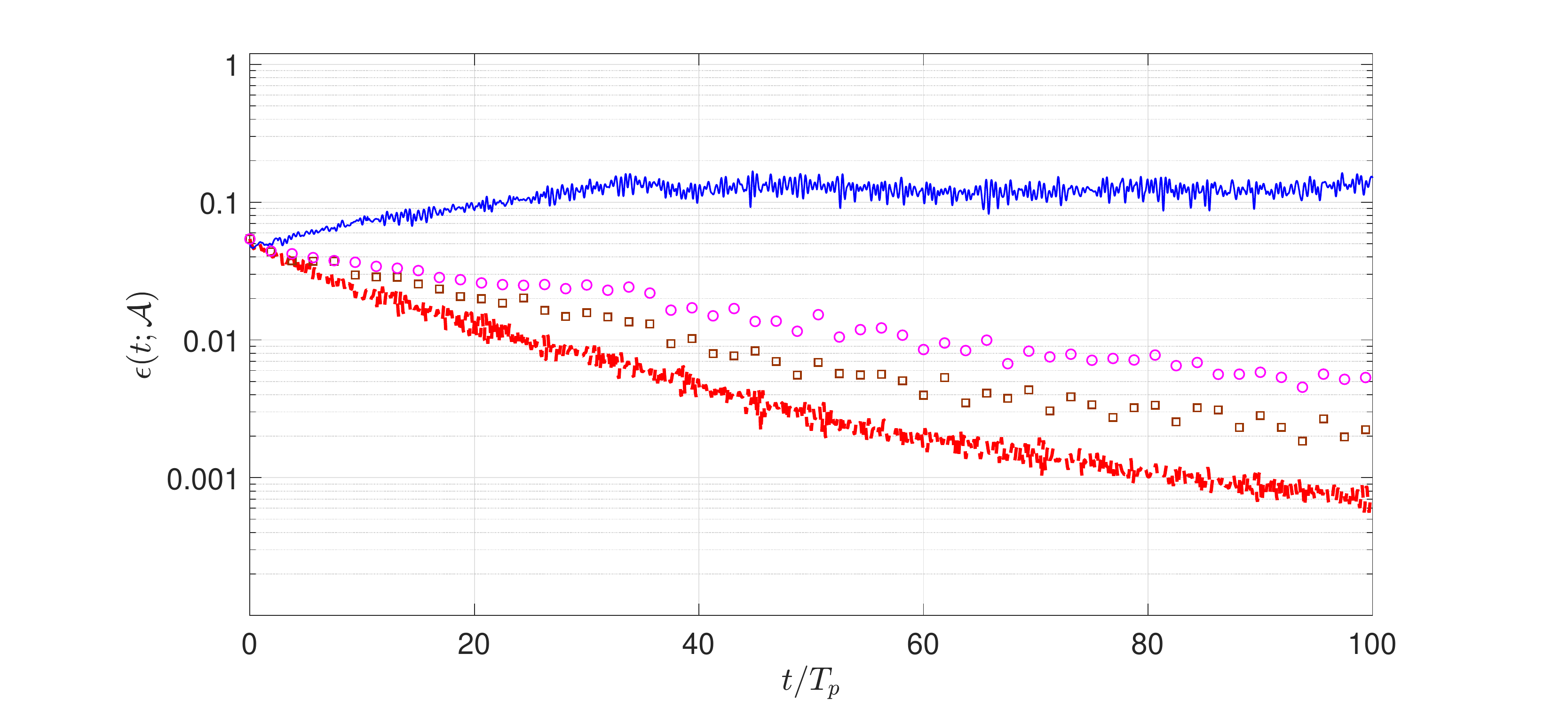}}
  \caption{Error $\epsilon(t;\mathcal{A})$ with $\mathcal{A}=[0,2\pi)$ from EnKF-HOS method for $\tau=T_p/16$ ({\color{red}\dashL}), $\tau=T_p/8$ ({\color{brown}$\Box$}), $\tau=T_p/4$ ({\color{magenta}$\circ$}) and $\tau=T_p/2$ ({\color{blue}\rule[0.5ex]{0.5cm}{0.25pt}}). Other parameter values are kept the same as the main 2D idealistic case.}
\label{fig:compare_dw}
\end{figure}

\begin{figure}
  \centerline{\includegraphics[scale =0.4]{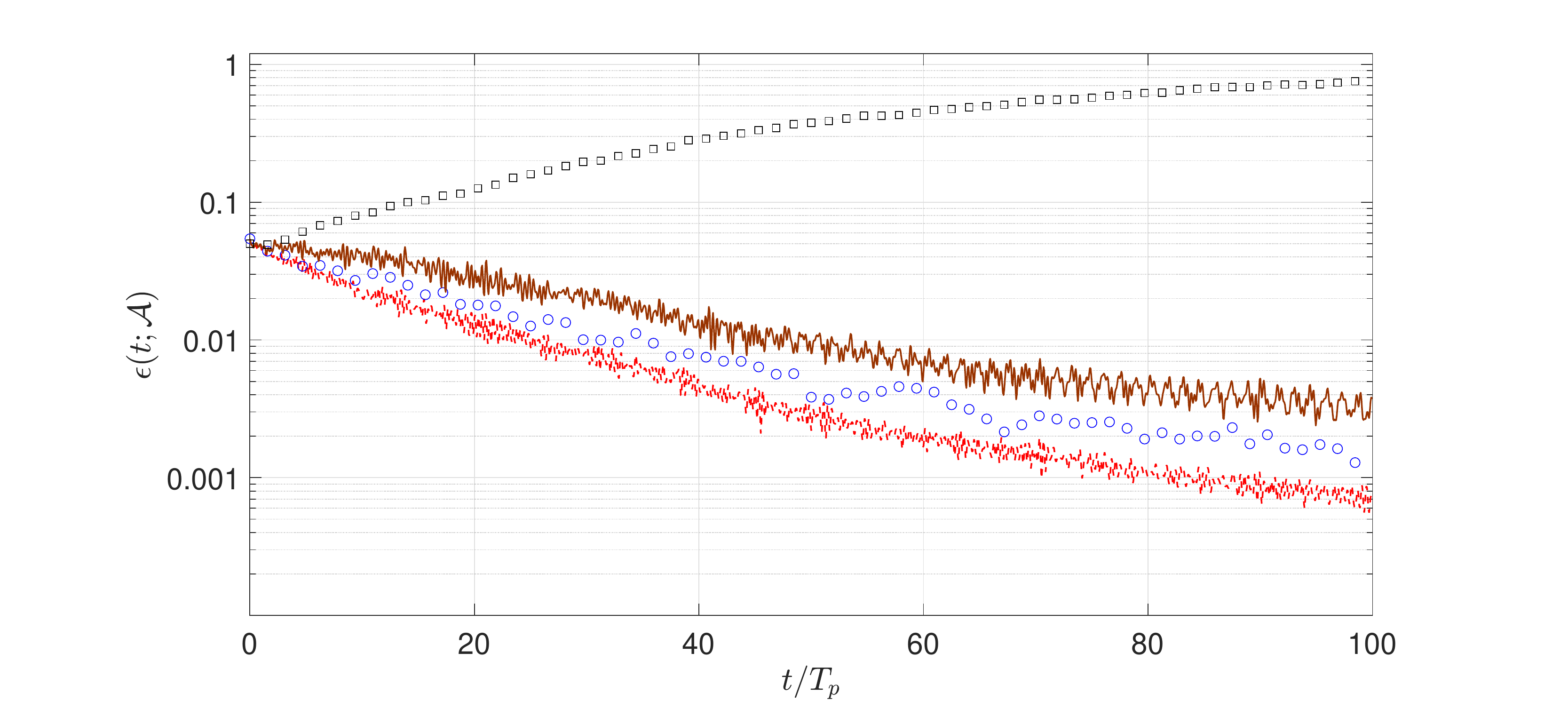}}
  \caption{Error $\epsilon(t;\mathcal{A})$ with $\mathcal{A}=[0,2\pi)$ from EnKF-HOS method for $N=20$ ($\Box$), $N=40$ ({\color{brown}\rule[0.5ex]{0.5cm}{0.25pt}}), $N=70$ ({\color{blue}$\circ$}), and $N=100$ ({\color{red}\dashL}). Other parameter values are kept the same as the main 2D idealistic case.}
\label{fig:compare_N}
\end{figure}

\begin{figure}
  \centerline{\includegraphics[scale =0.4]{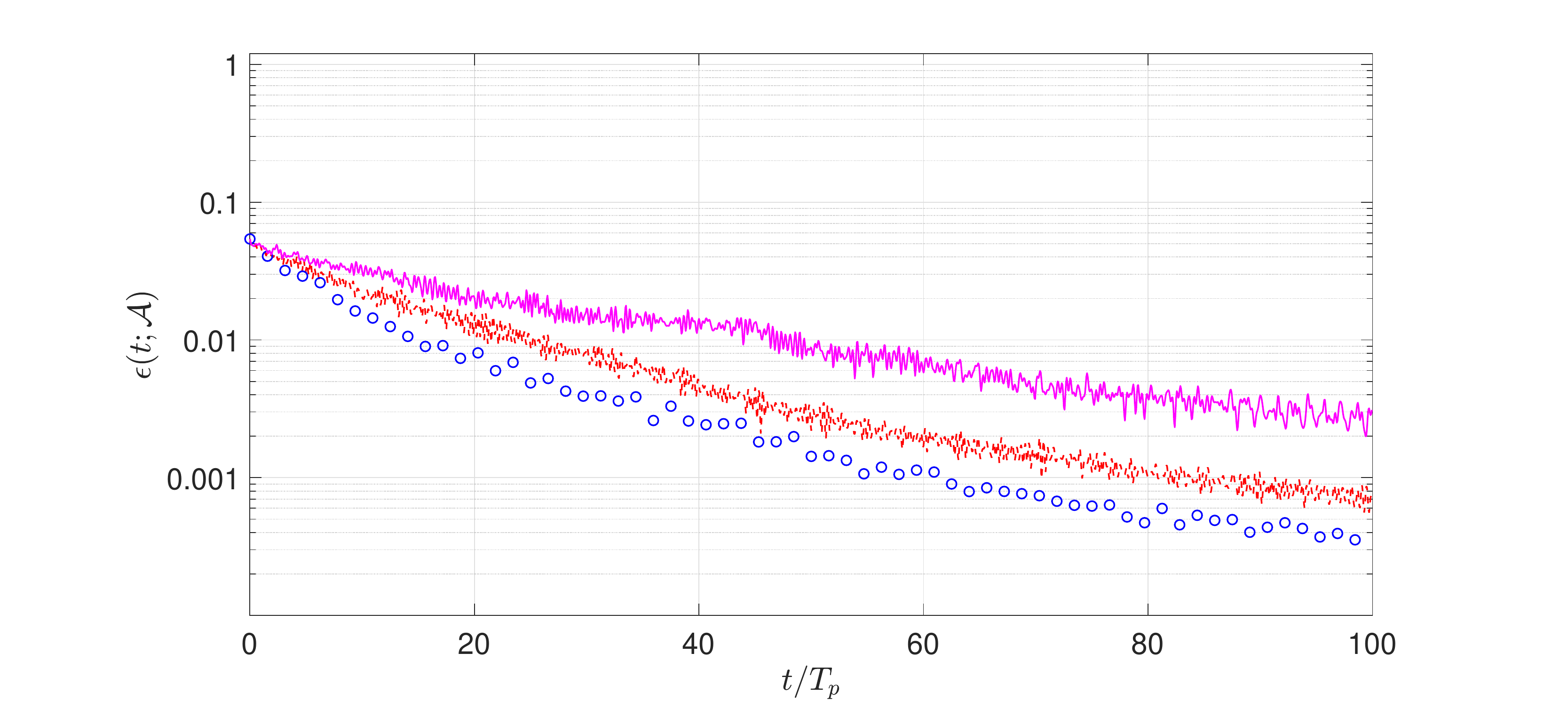}}
  \caption{Error $\epsilon(t;\mathcal{A})$ with $\mathcal{A}=[0,2\pi)$ from EnKF-HOS method for $d=1$ at $x/(2\pi)=100/256$ ({\color{magenta}\rule[0.5ex]{0.5cm}{0.25pt}}), $d=2$ at $x/(2\pi)=100/256~\text{and}~170/256$ ({\color{red}\dashL}) and $d=4$ at $x/(2\pi)=100/256,~135/256,~170/256,~\text{and}~205/256$ ({\color{blue}$\circ$}). Other parameter values are kept the same as the main 2D idealistic case.}
\label{fig:compare_d}
\end{figure}

We further investigate the effects of EnKF parameters on the performance, including DA interval~$\tau$, the ensemble size $N$, and the number of DA locations $d$. The errors $\epsilon(t;\mathcal{A})$ obtained with different parameter values are plotted in figure \ref{fig:compare_dw} (for $\tau$ from $T_p/16$ to $T_p/2$), figure \ref{fig:compare_N} (for $N$ from 40 to 100) and figure \ref{fig:compare_d} (for $d$ from 1 to 4). In the tested ranges, the performance of EnKF-HOS is generally better (i.e., faster decrease of $\epsilon(t;\mathcal{A})$ with increase of $t$) for smaller $\tau$, larger $N$ and larger $d$. In addition, for $\tau=T_p/2$ as shown in figure \ref{fig:compare_dw}, $\epsilon(t;\mathcal{A})$ slowly increases with time, indicating a situation that the assimilated data is not sufficient to counteract the deviation of HOS simulation from the true solution (due to the chaotic nature of \eqref{eq:bc1} and \eqref{eq:bc2}). It is also found that when $N=20$, the error increases rapidly leading to a filter divergence, mainly due to the insufficient ensemble size to capture the error statistics.

For the 3D wave field, we use the same initial spectrum $S(\omega)$ as in the 2D case, with a direction spreading function
\begin{equation}
    D(\theta)=
    \begin{cases}
    \displaystyle\frac{2}{\beta}\cos^2(\displaystyle\frac{\pi}{\beta}\theta),~&\text{for}~\displaystyle-\frac{\beta}{2}<\theta<\displaystyle\frac{\beta}{2} \\
    0,~&\text{otherwise}
    \end{cases}
\end{equation}
where $\beta=\pi/6$ is the spreading angle. The (reference, ENKF-HOS and HOS-only) simulations are conducted with $L=64\times64$ grid points. In EnKF-HOS method, $N=100$ ensemble members are used, and data from $d=10$ locations (randomly selected with uniform distribution) are assimilated with interval $\tau=T_p/16$.
\begin{figure}
  \centerline{\includegraphics[scale =0.4]{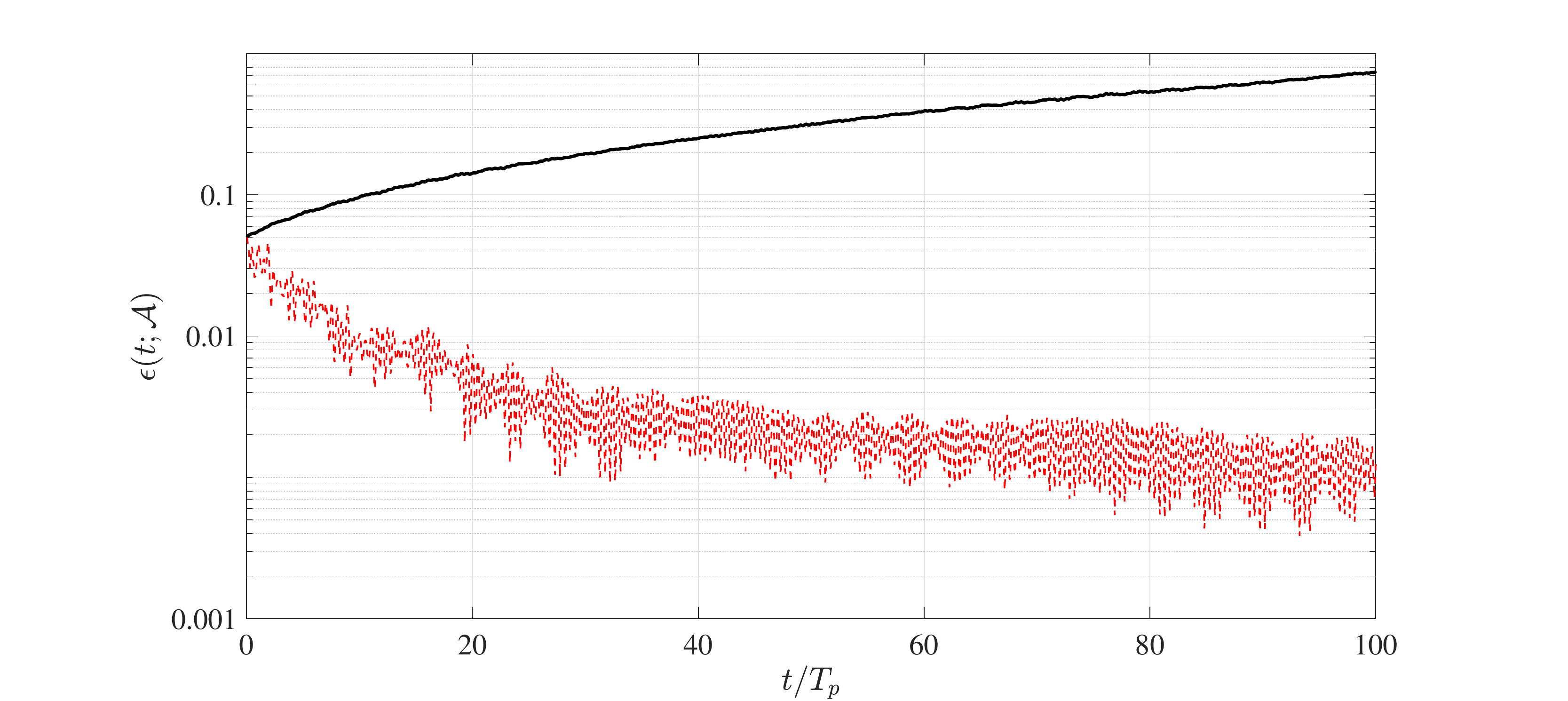}}
  \caption{Error $\epsilon(t;\mathcal{A})$ with $\mathcal{A}=[0,2\pi)\times[0,2\pi)$ from EnKF-HOS ({\color{red}\dashL}) and HOS-only (\rule[0.5ex]{0.5cm}{0.25pt}) methods, for the 3D idealistic case. }
\label{fig:compare_3d}
\end{figure}

\begin{figure}
  \centerline{\includegraphics[trim=0cm 0cm 0cm 0cm, clip,scale =0.4]{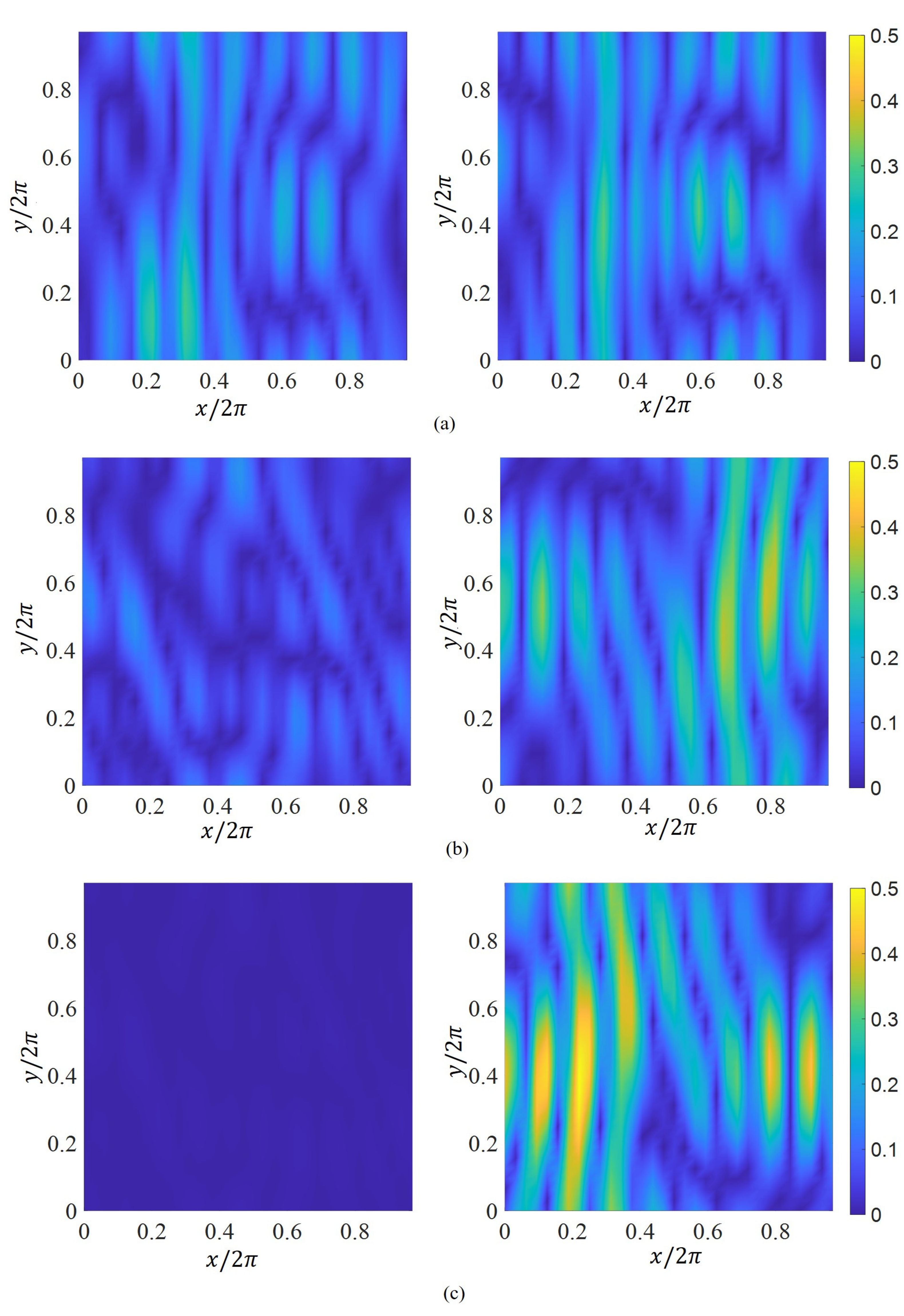}}
  \caption{Local spatial error $e(\boldsymbol{x};t)$ obtained with EnKF-HOS (left) and HOS-only (right) methods at (a) $t/T_p=5$, (b) $t/T_p=50$ and (c) $t/T_p=95$ for the 3D idealistic case.}
\label{fig:error_map_3d}
\end{figure}
Figure~\ref{fig:compare_3d} shows the error $\epsilon(t;\mathcal{A})$ (with $\mathcal{A}=[0,2\pi)\times[0,2\pi)$) obtained from the EnKF-HOS and HOS-only methods. Similar to the 2D case, we see that $\epsilon(t;\mathcal{A})$ from the EnKF-HOS method decreases with time, and becomes several orders of magnitude smaller than that from the HOS-only method (with the latter increasing with time). A closer scrutiny for error on a snapshot can be obtained by defining a local spatial error at a time instant $t$:  
\begin{equation}
e(\boldsymbol{x};t)=\frac{\mid\eta^{\text{true}}(\boldsymbol{x},t)-\eta^{\text{sim}}(\boldsymbol{x},t)\mid}{\sigma_{\eta}}.
\end{equation}
Three snapshots at $e(\boldsymbol{x};t)$ for $t/T_p=5,~50$ and $95$ are shown in figure \ref{fig:error_map_3d}, demonstrating the much smaller error achieved using EnKF-HOS method especially for large $t$, i.e., the superior performance of including DA in the simulation.

\subsubsection{Results for realistic cases}
\begin{figure}
  \centerline{\includegraphics[trim=0cm 0cm 0cm 0cm, clip,scale=0.3]{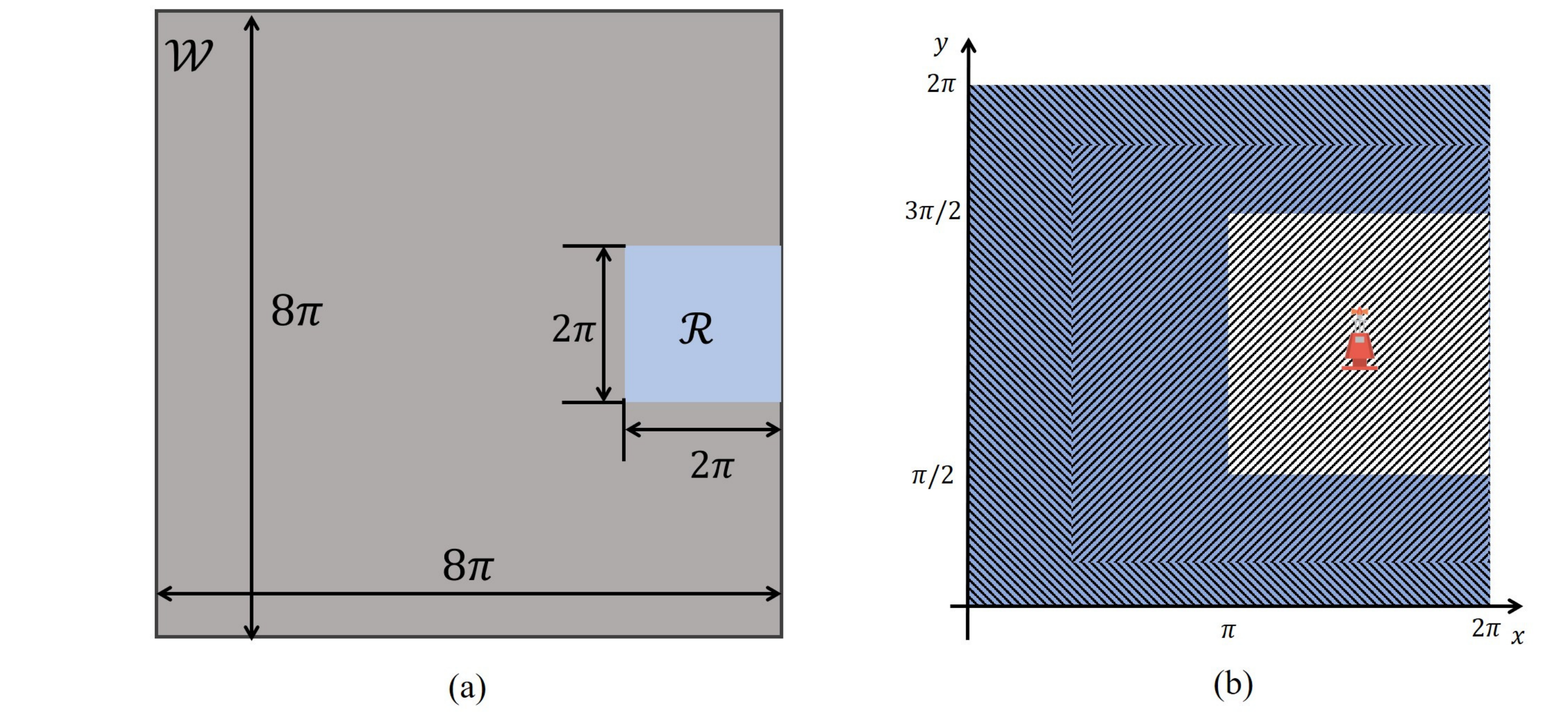}}
  \caption{Schematic illustration of spatial domains: (a) Periodic computational domain $\mathcal{W}$ for the reference simulation and sub-region $\mathcal{R}$ for EnKF-HOS and HOS-only simulations; (b) Computational region $\mathcal{R}$ with measurement zone $\mathcal{M}_j$ marked by blue, predictable zone $\mathcal{P}_j$ by upward diagonal stripes, and unpredictable zone $\mathcal{U}_j$ by downward diagonal stripes.}
\label{fig:pzone3d}
\end{figure}

We consider $\eta^{\text{true}}(\boldsymbol{x})$ for the realistic case taken from a sub-region $\mathcal{R}$ with quarter edge length of a periodic computational domain $\mathcal{W}$, i.e., a patch in the ocean (see figure \ref{fig:pzone3d}(a)). The reference simulation in $\mathcal{W}$ is performed with $256\times 256$ grid points, with all other parameters kept the same as the 3D idealistic reference simulation. The EnKF-HOS and HOS-only simulations are conducted over $\mathcal{R}=[0,2\pi)\times[0,2\pi)$ with $L=64\times64$ grid points, starting from initial noisy measurements. For $j\geq1$, We further consider a practical situation where the measurements are only available in $\mathcal{M}_j=\mathcal{B}^{\text{c}}\cap\mathcal{R}$, where $\mathcal{B}=\{\boldsymbol{x}|x>\pi, \pi/2<y<3\pi/2\}$ (say a structure of interest located within $\mathcal{B}$ preventing the surrounding measurements, see figure \ref{fig:pzone3d}(b)). We use $d=2176$ (locating on every computational grid point in $\mathcal{M}_j$) and an assimilation interval $\tau=T_p/4$.

In this case, the use of modified EnKF analysis equation \eqref{eq:ananew} is critical due to the interplay among $\mathcal{M}_j$, $\mathcal{P}_j$ and $\mathcal{U}_j$. In particular, the left, upper/lower bounds of $\mathcal{P}(t)$ moves (towards right, down/up) respectively with speeds $c_{g,x}^{\text{max}}$ and $c_{g,y}^{\text{max}}$, i.e., the maximum group speeds in $x$ and $y$ directions (practically taken as group speed of mode $\boldsymbol{k}=(4,1)$, so that the energy from longer waves is less than $1\%$ of the total energy). After applying the modified EnKF analysis equation~\eqref{eq:ananew}, which takes into consideration of $\mathcal{M}_j$, $\mathcal{P}_j$ and $\mathcal{U}_j$ (see a sketch in figure \ref{fig:pzone3d}(b)), $\mathcal{P}_j$ is recovered to fill in $\mathcal{R}$ due to the DA. In contrast, in HOS-only method, $\mathcal{P}(t)\equiv \mathcal{P}^*(t)$ keeps shrinking and vanishes for sufficient time.

\begin{figure}
  \centerline{\includegraphics[trim=1cm 0cm 0cm 0cm, clip,scale=0.4]{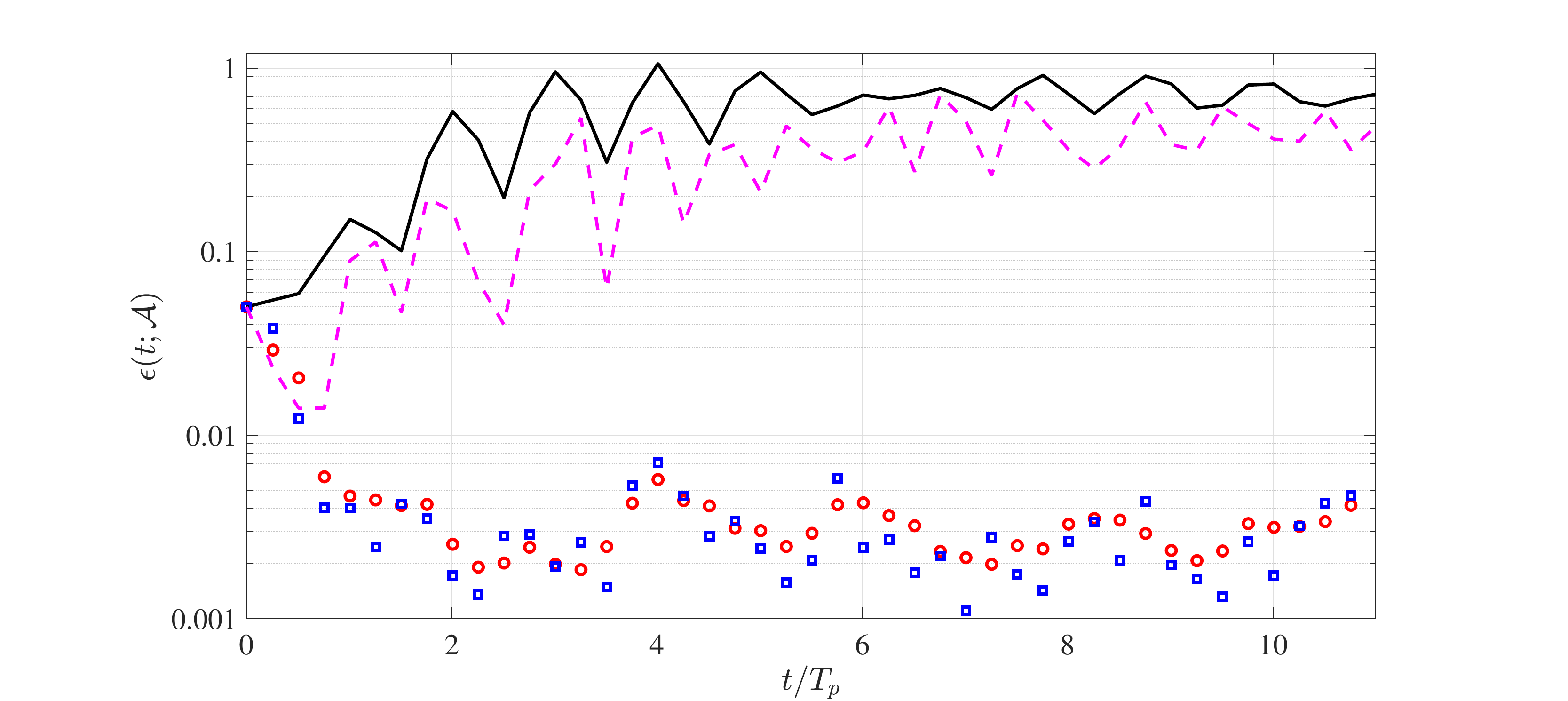}}
  \caption{The error metric $\epsilon(t; \mathcal{R})$ obtained from EnKF-HOS ({\color{red}$\circ$}) and HOS-only (\rule[0.5ex]{0.5cm}{0.25pt}) methods; and $\epsilon(t; \mathcal{P}^*(t))$ obtained from EnKF-HOS ({\color{blue}$\Box$}) and HOS-only ({\color{magenta}\dashL}) methods, for the 3D realistic case.}
\label{fig:practical_error}
\end{figure}

We consider two error metrics $\epsilon(t; \mathcal{R})$ and $\epsilon(t; \mathcal{P}^*(t))$, which are plotted in figure \ref{fig:practical_error} for both EnKF-HOS and HOS-only methods. For HOS-only simulation, $\epsilon(t; \mathcal{R})$ increases rapidly in time and reaches $\mathcal{O}(1)$ at $t/T_p\sim \mathcal{O}(3)$. This is resulted from the chaotic nature of \eqref{eq:bc1} and \eqref{eq:bc2}, as well as the significant error in $\mathcal{U}(t)$. For EnKF-HOS simulation, $\epsilon(t; \mathcal{R})$ decreases with time and reaches a constant level of $\mathcal{O}(0.002)$ after $t/T_p\sim \mathcal{O}(3)$. The further reduction of the error is prohibited due to the region $\mathcal{U}(t)$, because $\epsilon(t; \mathcal{U}(t))$ has a lower bound from the measurement error. The general trend of $\epsilon(t; \mathcal{P}^*(t))$ is similar for both methods, but the magnitude of $\epsilon(t; \mathcal{P}^*(t))$ is smaller than that of $\epsilon(t; \mathcal{R})$ especially for the HOS-only method. This is due to the removal of $\mathcal{U}(t)$ from $\mathcal{A}$ in the computation of the error.

\begin{figure}
  \centerline{\includegraphics[trim=0cm 0cm 0cm 0cm, clip,scale =0.4]{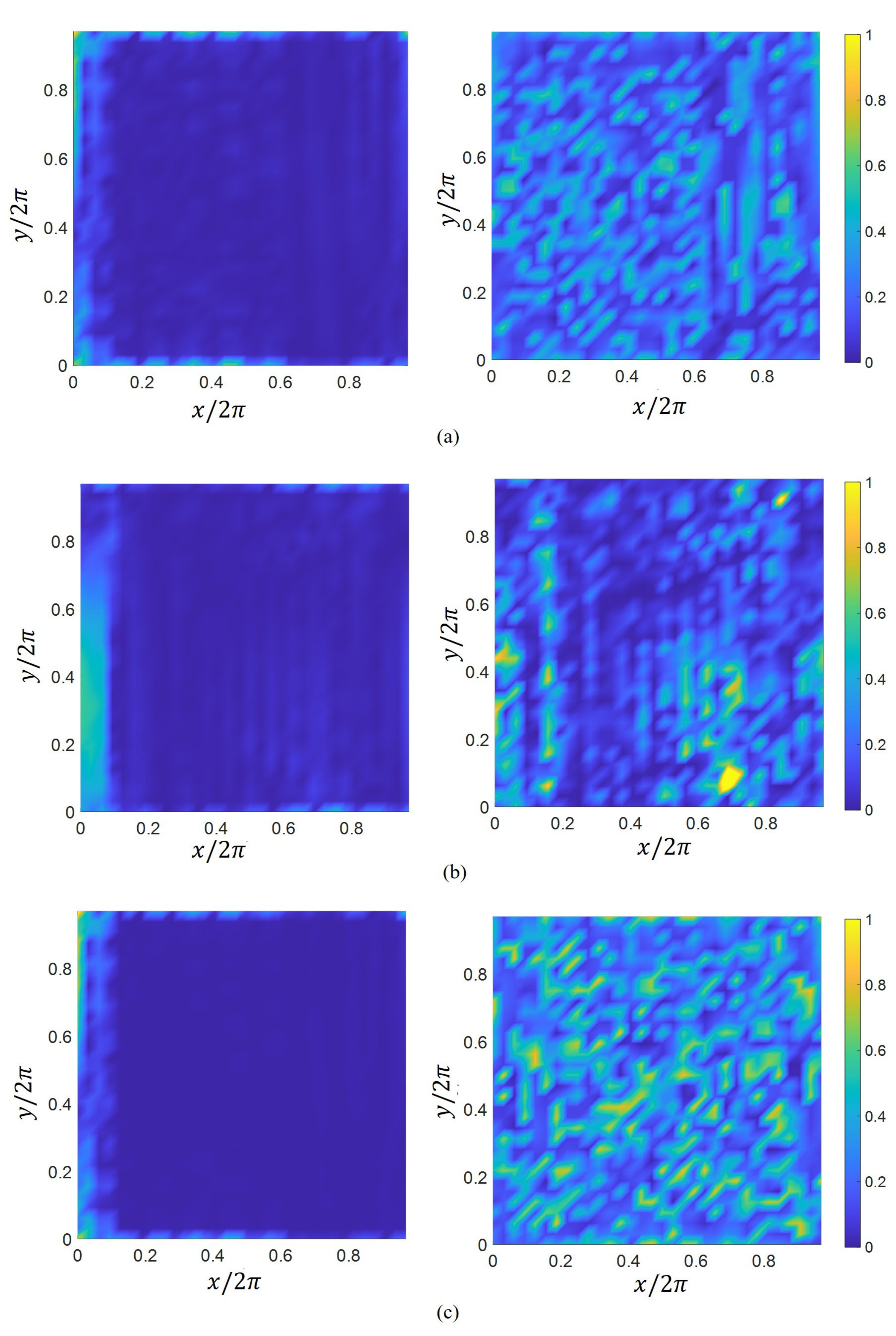}}
  \caption{Local spatial error $e(\boldsymbol{x};t)$ obtained from (left) EnKF-HOS and (right) HOS-only methods for the 3D realistic case, at (a) $t/T_p=2$, (b) $t/T_p=5$ and (c) $t/T_p=8$.}
\label{fig:error_map_3d_prac}
\end{figure}

To further understand the error characteristics, we plot the spatial local error $e(\boldsymbol{x};t)$ at $t/T_p=2,~ 5,$ and $8$ for both methods in figure \ref{fig:error_map_3d_prac}. While the spatial error generally increases in time for HOS-only method, $e(\boldsymbol{x};t)$ from EnKF-HOS method is significantly smaller and exhibits heterogeneous spatial distribution. Within $\mathcal{U}_j$, $e(\boldsymbol{x};t)$ is relatively high with the same order of the measurement error. In $\mathcal{P}_j$, $e(\boldsymbol{x};t)$ decreases with time and becomes significantly lower than that in $\mathcal{U}_j$. Remarkably, this also applies to the region where measurements are not available (i.e., $\mathcal{M}^c\cap\mathcal{R}$) as the waves in this region travels from upstream locations where DA is performed. This result is of practical importance as it shows that the wave forecast at a location of interest in the ocean (say the location of an offshore structure) can be made accurate through DA in the upstream region. 

\subsection{Results with real radar measurements}
\begin{figure}
  \centerline{\includegraphics[trim=0cm 0cm 0cm 0cm, clip,scale=0.7]{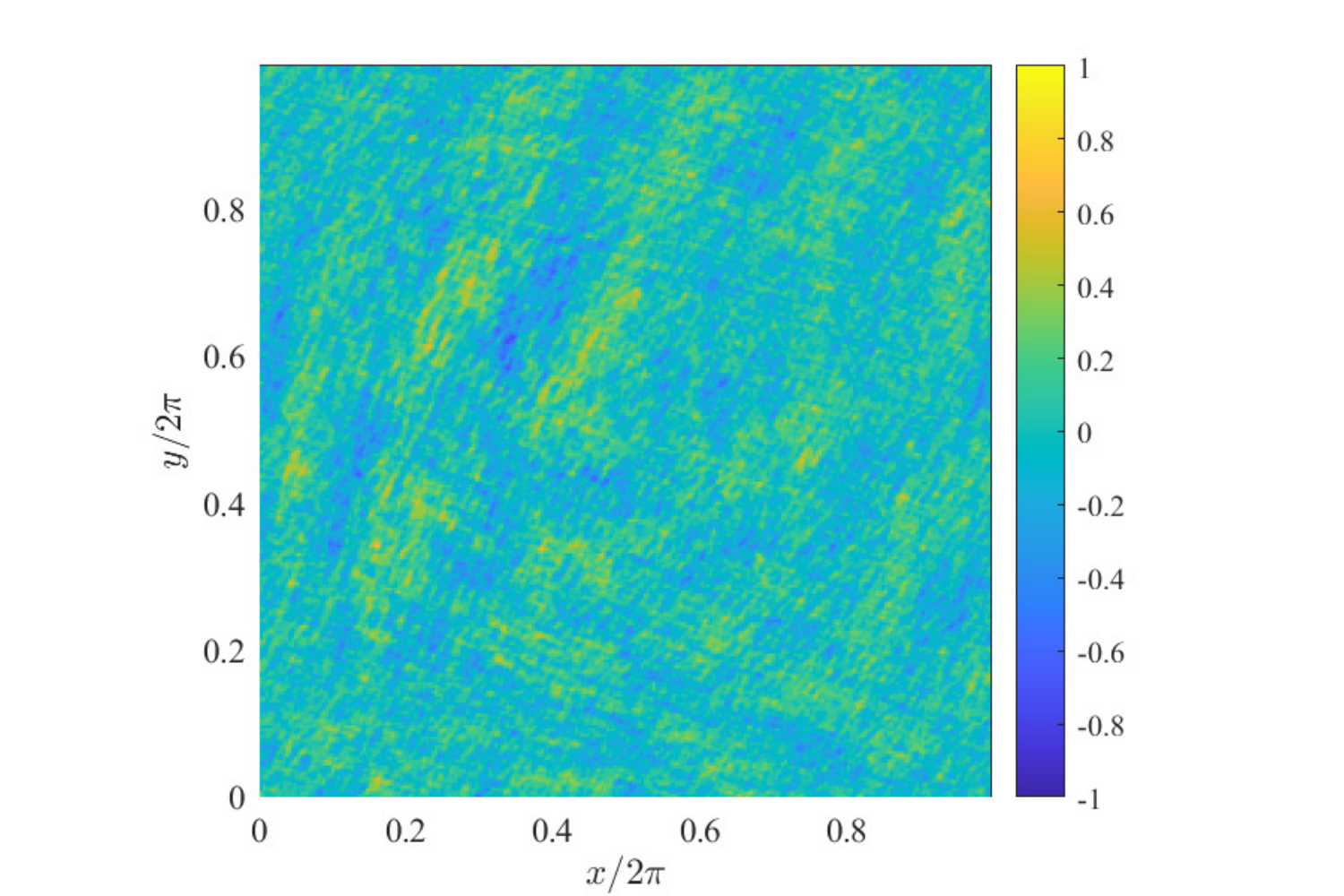}}
  \caption{Initial surface elevation $\eta_{m,0}(\boldsymbol{x})/H_s$ (with $H_s=1.70\text{m}$) measured by radar at $t=t_0$, i.e., $2013$-$09$-$13\text{T}01:00:13\text{Z}$.}
\label{fig:ic_eta_real}
\end{figure}

In this section, we test the performance of EnKF-HOS method with real radar measurements of the ocean wave field. The measurements are obtained from a ship-borne X-band ($9.4~\text{GHz}$) Doppler coherent marine radar off the coast of southern California. We consider a patch from the radar-scanned area as our initial domain of interest $\mathcal{R}_0$, which covers a 480m$\times$480m area resolved on a $64\times64$ grid (Figure~\ref{fig:ic_eta_real}). The starting time of computation is $2013$-$09$-$13\text{T}01:00:13\text{Z}$, with a global wave steepness $k_pH_s/2=0.027$ from the initial radar data. In both EnKF-HOS and HOS-only simulations, we re-scale the computational domain $\mathcal{R}_j$ to $[0,2\pi)\times[0,2\pi)$ and use $L=64\times64$ grid points. For EnKF, we use $d=64\times64$, which covers the whole patch, and set the DA interval the same as radar data collection interval, which fluctuates in time around $T_p/4=2.82\text{s}$. The ensemble size is set to be $N=100$. For adaptive inflation, we use $\bar{\lambda}_0=1$ in the prior distribution of $\lambda_0$ (see Appendix A) to sequentially determine $\lambda_j$ in \eqref{eq:inflationeta}, which is applied at each $t=t_j$ together with the localization \eqref{eq:loc}. 

A critical issue in this case is the movement of $\mathcal{M}_j$ in time due to the ship speed, which results in a mismatch between $\mathcal{M}_j$ and the computational region $\mathcal{R}_{j-1}$ at each $t=t_j$. To address this issue, we shift the computational region from $\mathcal{R}_{j-1}$ to $\mathcal{R}_j$ which matches $\mathcal{M}_j$. In EnKF-HOS method, we further partition $\mathcal{R}_j$ into $\mathcal{P}_j$ and  $\mathcal{U}_j$ (using the predictable zone calculated from $\mathcal{R}_{j-1}$) and apply the modified analysis equation \eqref{eq:ananew} accordingly. In HOS-only method, we use Fourier periodic extension~\citep{grafakos2008classical} to obtain the wave field covering $\mathcal{R}_j$.

\begin{figure}
  \centerline{\includegraphics[trim=0cm 0cm 0cm 0cm, clip,scale=0.35]{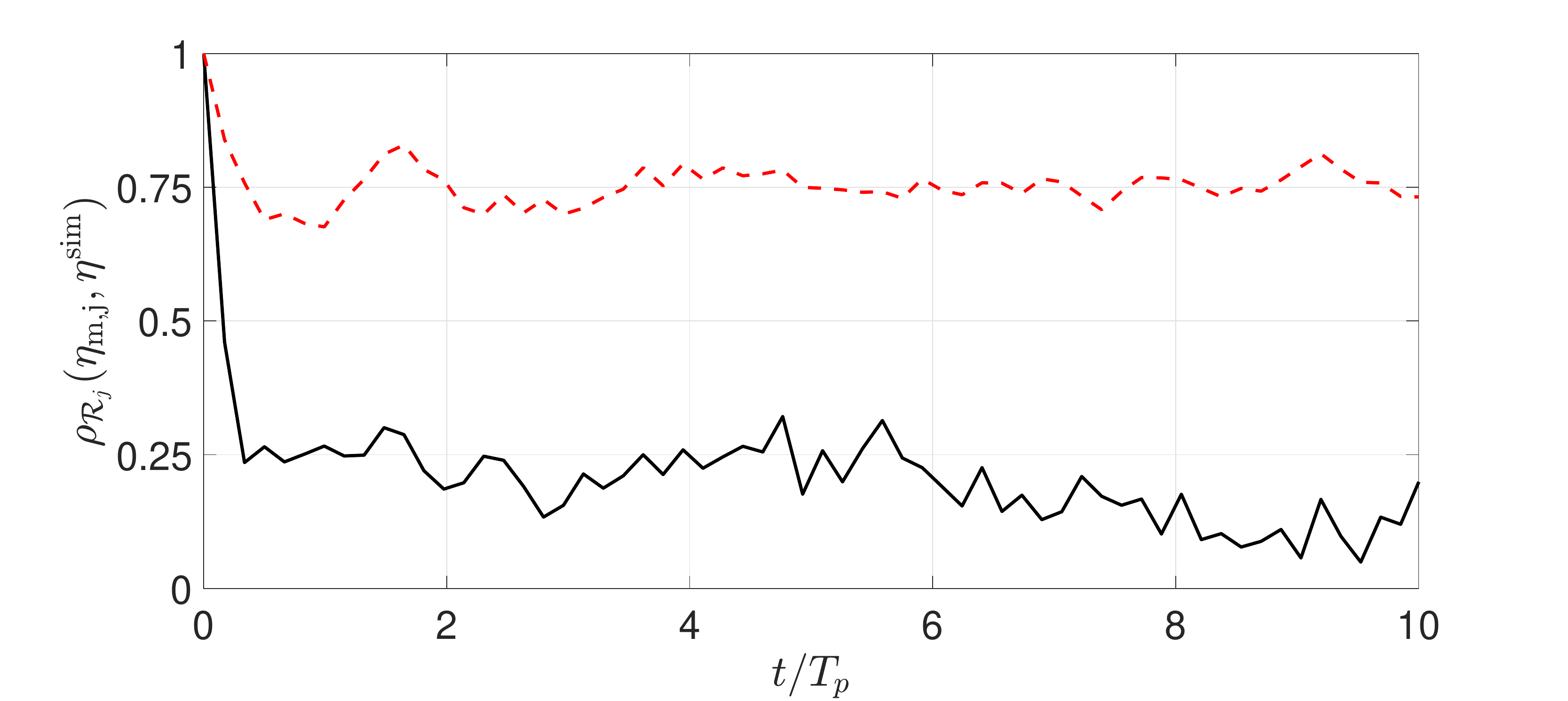}}
  \caption{Time series of $\rho_{\mathcal{R}_j}(\eta_{\text{m},j}, \eta^{\text{sim}})$ from EnKF-HOS ({\color{red}\dashL}) and HOS-only (\rule[0.5ex]{0.5cm}{0.25pt}) methods.}
\label{fig:real_error}
\end{figure}
\begin{figure}
  \centerline{\includegraphics[trim=0cm 0cm 0cm 0cm, clip,scale =0.4]{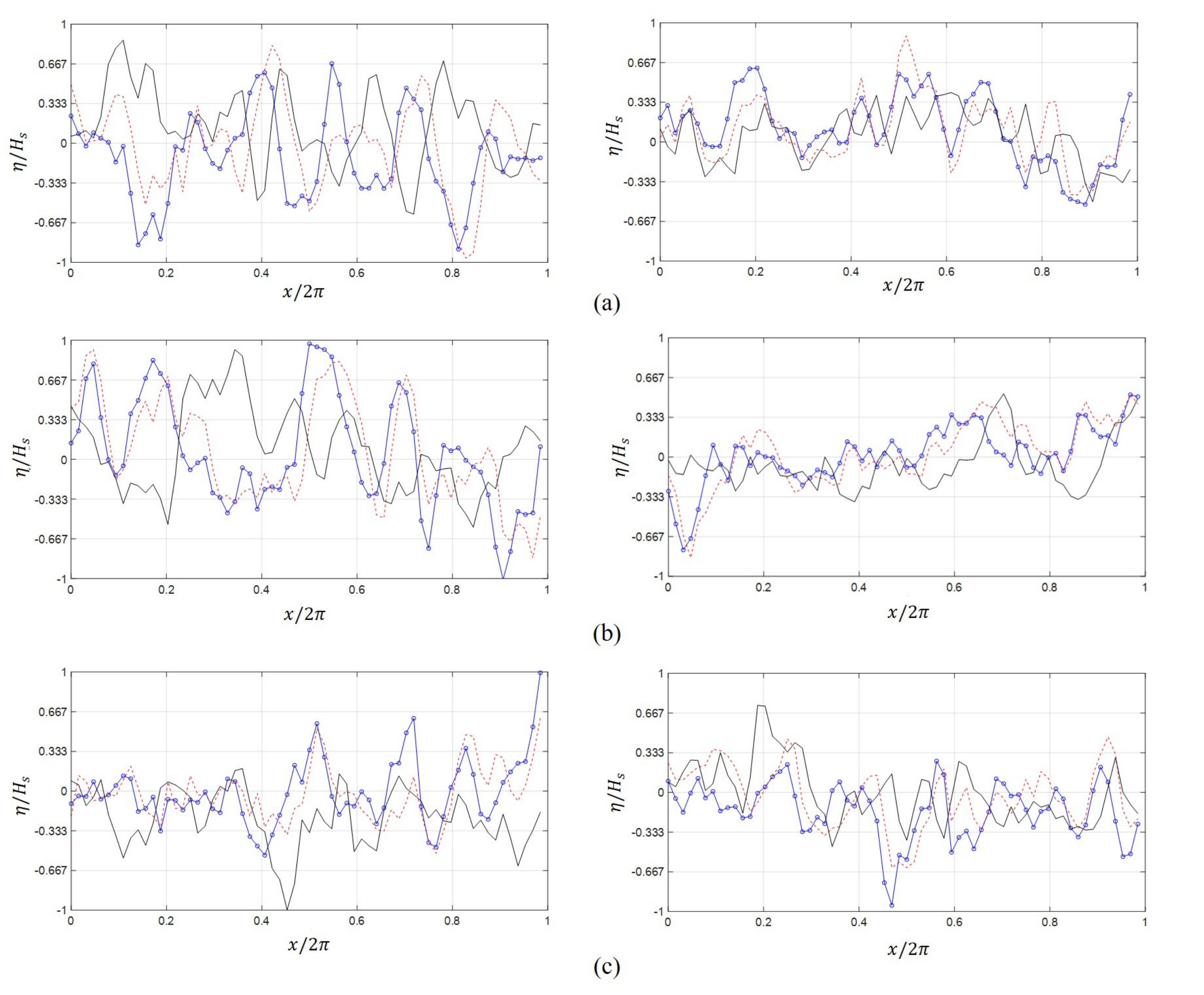}}
  \caption{Snapshots of $\eta_{\text{m},j}(\boldsymbol{x})$ ({\color{blue}\Lcirc}) and $\eta^{\text{sim}}(\boldsymbol{x},t)$ from EnKF-HOS ({\color{red}\dashL}) and HOS-only (\rule[0.5ex]{0.5cm}{0.25pt}) methods, at two cross-sections (left) $y/(2\pi)=1/3$ and (right) $y/(2\pi)=2/3$, and time instants (a) $t/T_p=1$, (b) $t/T_p=5$ and (c) $t/T_p=9$.}
\label{fig:compare_snap_3d_real}
\end{figure}

Since the true solution is not available in this case, we directly use $\rho_{\mathcal{R}_j}(\eta_{\text{m},j}, \eta^{\text{sim}})$ (see \eqref{eq:corr}) as the metric to evaluate the performance. Figure \ref{fig:real_error} plots $\rho_{\mathcal{R}_j}(\eta_{\text{m},j}, \eta^{\text{sim}})$ obtained from EnKF-HOS and HOS-only methods. While both time series starts from $\rho_{\mathcal{R}_0}=1$ at $t=t_0$, the one from HOS-only simulation quickly approaches $\mathcal{O}(0.25)$ within one peak period $T_p$, indicating the (almost) complete loss of the phase information. This is much faster than any synthetic case, mainly due to the under-resolved physics in \eqref{eq:bc1} and \eqref{eq:bc2} with respect to the real ocean (which includes extra physical effects of current, wind, etc.). In contrast, the correlation $\rho_{\mathcal{R}_j}(\eta_{\text{m},j}, \eta^{\text{sim}})$ from EnKF-HOS remains at $\mathcal{O}(0.75)$, retaining the phase information for arbitrarily long time. Figure~\ref{fig:compare_snap_3d_real} further plots the snapshots of $\eta_{\text{m},j}(\boldsymbol{x})$ and $\eta^{\text{sim}}(\boldsymbol{x},t)$ at two cross-sections of $y/(2\pi)=1/3$ and $2/3$ for both methods at three time instants of $t/T_p=1, 5$ and $9$. It can be visually observed that the EnKF-HOS results are (on average) much closer to $\eta_{\text{m},j}(\boldsymbol{x})$ for all cases.

\begin{figure}
  \centerline{\includegraphics[trim=0cm 0cm 0cm 0cm, clip,scale=0.35]{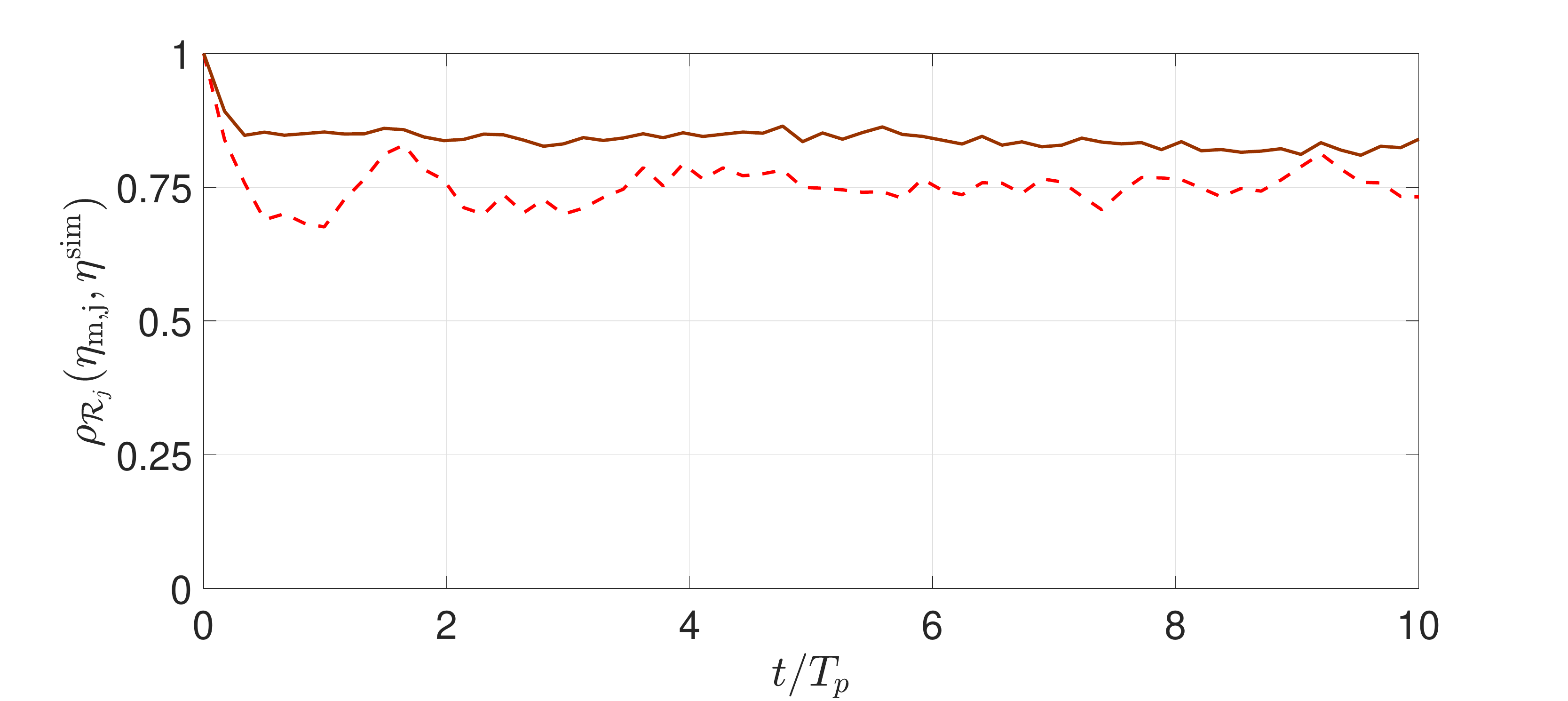}}
  \caption{Time series of $\rho_{\mathcal{R}_j}(\eta_{\text{m},j}, \eta^{\text{sim}})$ with $\bar\lambda_0=1.0$ (\dashL) and $\bar\lambda_0=2.0$({\color{brown}\rule[0.5ex]{0.5cm}{0.25pt}})}
\label{fig:real_error_lambda}
\end{figure}

We finally test the effect of parameter $\bar{\lambda}_0$ on the performance of EnKF-HOS. In general, the value of $\bar{\lambda}_0$ can be considered as a control of the extent to which the inflation is applied. For larger value of $\bar{\lambda}_0$, it is expected that the ensemble variance of \eqref{eq:etaen} is amplified to a greater extent, and more weights are assigned to measurements when the analysis \eqref{eq:ananew} is applied. Physically, larger values of $\bar{\lambda}_0$ may be chosen if the model is associated with significant under-represented physics (thus severely underestimates the variance in \eqref{eq:etaen}). To elucidate this effect of $\bar{\lambda}_0$, we test another value of $\bar{\lambda}_0=2$, and compare the resulting $\rho_{\mathcal{R}_j}(\eta_{\text{m},j}, \eta^{\text{sim}})$ with that from $\bar{\lambda}_0=1$ in figure \ref{fig:real_error_lambda}. Indeed, the result for $\bar{\lambda}_0=2$ shows a higher correlation with the measurement, with $\rho_{\mathcal{R}_j}$ above $\mathcal{O}(0.8)$ for all time. We note that the higher value of $\rho_{\mathcal{R}_j}$ does not imply the closeness of $\eta^{\text{sim}}$ to $\eta^{\text{true}}$, since the relation between $\eta_{\text{m},j}$ and $\eta^{\text{true}}$ is not known in this case.

\section{Conclusions}
\label{sec:conc}
In this paper, we develop the ensemble-based DA capability for phase-resolved wave forecast, resulting in a new EnKF-HOS method. A unique consideration in EnKF-HOS is the treatment of the interplay between predictable and measurement zones, which is successfully accounted for through a modified analysis equation. The performance of EnKF-HOS method is extensively tested and compared to the (traditional) HOS-only method using both synthetic wave field and real radar data. In all cases, significant advantages are demonstrated by using the EnKF-HOS method, namely the dramatic reduction of forecast error and retaining the phase information for arbitrarily long time with DA of radar data. In contrast, the phase information is lost within one peak period in HOS-only method when considering the real ocean waves and using the radar data. The parameters involved in the EnKF-HOS method is carefully benchmarked, including the ensemble size, DA interval, number of DA locations and the inflation factor. The developed EnKF-HOS algorithm is intrinsically parallel and very suitable to be implemented on a graphics processing unit (GPU), a compact device that can be conveniently installed in the offshore environment.

\section*{Acknowledgement}
We would like to thank Dr. David Lyzenga, Dr. Okey Nwogu and Dr. Robert Beck in the Department of Naval Architecture and Marine Engineering at the University of Michigan, for providing and interpreting the radar data. We also thank Dr. Paul Hess from the Office of Naval Research to grant us permission to use this data set for research. The research is funded by the Michigan Institute for Computational Discovery and Engineering (MICDE) through a 2019 Catalyst Grant. The computational resources are provided by NSF-XSEDE grant TG-CTS200016. 

\appendix
\section{Adaptive inflation}\label{appA}
We apply the adaptive inflation algorithm developed by~\cite{anderson2007adaptive} to determine the values of $\lambda_j$ in \eqref{eq:inflationeta}. The algorithm is applied at each $t=t_j$ (for $j\geq 1$), and we shall drop the subscript $j$ on other variables in the following description for simplicity.   

The key idea in adaptive inflation is to consider $\lambda$ as an additional state variable and update its value through Bayes' theorem given the measurements:
\begin{equation}
    p\big(\lambda|\eta_{\text{m}}(\boldsymbol{x})\big)\sim p(\lambda)p(\eta_{\text{m}}(\boldsymbol{x})|\lambda),
    \label{fullB}
\end{equation}
where $p(\lambda)$ is the (given) prior distribution, and $p(\eta_{\text{m}}(\boldsymbol{x})|\lambda)$ is the likelihood function. In principle, our purpose is to find the optimal $\lambda$ which maximizes the posterior $p\big(\lambda|\eta_{\text{m}}(\boldsymbol{x})\big)=p\big(\lambda|\eta_{\text{m}}(1),...,\eta_{\text{m}}(d)\big)$. Intuitively, the inflation \eqref{eq:inflationeta} using such an optimal $\lambda$ provides sufficient variance of the forecast ensemble to cover the measurements, and thus avoid the overconfidence in the forecast when analysis is performed.

By assuming independent measurement errors, it can be shown that the full Bayes' problem \eqref{fullB} is equivalent to the sequential problem (which saves significant computational cost, see \cite{anderson2007adaptive} for details): 
\begin{equation}
    p\big(\lambda|\eta_{\text{m}}(1)\big)=p\big(\eta_{\text{m}}(1)|\lambda\big)p\big(\lambda\big)/\mathcal{Z}_1,
\label{eq:lam1}
\end{equation}
\begin{equation}
    p\big(\lambda|\eta_{\text{m}}(1),\eta_{\text{m}}(2),...,\eta_{\text{m}}(i)\big)=p\big(\eta_{\text{m}}(i)|\lambda\big)p\big(\lambda|\eta_{\text{m}}(1),\eta_{\text{m}}(2),...,\eta_{\text{m}}(i-1)\big)/\mathcal{Z}_i,
\label{eq:lam2}
\end{equation}
where \eqref{eq:lam2} is applied sequentially for $i=2,...,d$, and $\mathcal{Z}_i$ are the normalization factors (which do not play a role in the computation). 

In computing \eqref{eq:lam1} and \eqref{eq:lam2}, we use a Gaussian prior distribution:
\begin{equation}
    p(\lambda)=\mathcal{N}(\bar{\lambda}_0,\sigma_0^2),
\end{equation}
with mean $\bar{\lambda}_0$ and variance $\sigma^2_0$ (say $\bar{\lambda}_0=1$ and $\sigma^2_0={c\bar{\lambda}_0^2}/{H_s^2}$ at $j=1$, with the situation of $j>1$ discussed at the end of Appendix A). 
The key computations in \eqref{eq:lam1} and \eqref{eq:lam2} are the likelihood functions $p\big(\eta_{\text{m}}(i)|\lambda\big)$ for $i=1,2,...,d$, which will be discussed below. 

Given a value of $\lambda$, a member in the forecast ensemble at measurement location $i$ is denoted by $[\boldsymbol{G}\eta_\text{f}^{(n),inf}](i)$ with $\eta_\text{f}^{(n),inf}$ given by \eqref{eq:inflationeta}. This ensemble is assumed to have a Gaussian distribution in consistency with the EnKF framework, with mean $[\boldsymbol{G}\bar{\eta}_{\text{f}}](i)$ and variance $\sigma_f(i)^2=\lambda [\boldsymbol{G}\boldsymbol{Q}_{\eta}\boldsymbol{G}^{\text{T}}](i,i)$ (note that only $\sigma_f(i)^2$ is affected by the inflation). Let $D=\eta_\text{m}(i) -[\boldsymbol{G}\bar{\eta}_{\text{f}}](i)$ be the distance between the mean forecast and measurement at location $i$, which is drawn from a zero-mean Gaussian random variable $\mathcal{D}$ with variance $\theta_i^2=\sigma_f(i)^2+\boldsymbol{R}_\eta(i,i)$ (here we consider the summation of two independent Gaussian random variables and assume that both the measurement and forecast are unbiased). It follows that    
\begin{equation}
p\big(\eta_\text{m}(i)|\lambda\big)=p\big(\mathcal{D}=D|\lambda\big)=\displaystyle\frac{1}{\sqrt{2\pi}\theta_i}\text{exp}\left(\displaystyle\frac{-D^2}{2\theta_i^2}\right).
\end{equation}
Therefore, the posterior distribution in each equation of \eqref{eq:lam1} and \eqref{eq:lam2} can be formulated as
\begin{eqnarray}
    p\big(\lambda|\eta_\text{m}(1),...,\eta_\text{m}(i)\big)=\displaystyle\frac{1}{{2\pi}\theta_i\sigma_{i-1}}\text{exp}\left(-\displaystyle\frac{D^2}{2\theta_i^2} -\frac{(\lambda-\bar{\lambda}_{i-1})^2}{2\sigma_{i-1}^2}\right)/\mathcal{Z}_i.
\label{eq:postlambda}
\end{eqnarray}
The formulation (and sequential computation) of \eqref{eq:postlambda} for $i$=$1,2,...,d$ require each posterior $p\big(\lambda|\eta_\text{m}(1),...,\eta_\text{m}(i)\big)$ in \eqref{eq:lam1} and \eqref{eq:lam2} (and thus the prior for the next sequential equation) to be approximated by Gaussian distribution $\mathcal{G}(\lambda)\sim \mathcal{N}(\bar{\lambda}_i,\sigma_i^2)$. We follow \cite{anderson2007adaptive} to set the $\bar{\lambda}_i$ as the mode of $p\big(\lambda|\eta_\text{m}(1),...,\eta_\text{m}(i)\big)$, and compute $\sigma_i$ by considering $\Gamma=p\big(\bar{\lambda}_i|\eta_\text{m}(1),...,\eta_\text{m}(i)\big)/p\big(\bar{\lambda}_i+\sigma_i|\eta_\text{m}(1),...,\eta_\text{m}(i)\big)=\mathcal{G}(\bar{\lambda}_i)/\mathcal{G}(\bar{\lambda}_i+\sigma_i)$, i.e., the same decay rate of the distribution, which gives $\sigma^2_i=-(\sigma^2_{i-1}/2)\ln{\Gamma}$. 

Finally, we use $\lambda=\bar{\lambda}_d$ in \eqref{eq:inflationeta} for inflation at $t=t_j$, and set $\mathcal{N}(\bar{\lambda}_d,\sigma_d^2)$ computed at time $t=t_j$ as the prior $p(\lambda)$ at $t=t_{j+1}$. The computations of \eqref{eq:lam1} and \eqref{eq:lam2} are repeated, which completes the full algorithm to determine $\lambda$ by adaptive inflation at each $t_j$.

\section{Gaspari-Cohn function}\label{appB}
\begin{figure}
  \centerline{\includegraphics[trim=0cm 0cm 0cm 0cm, clip,scale=0.45]{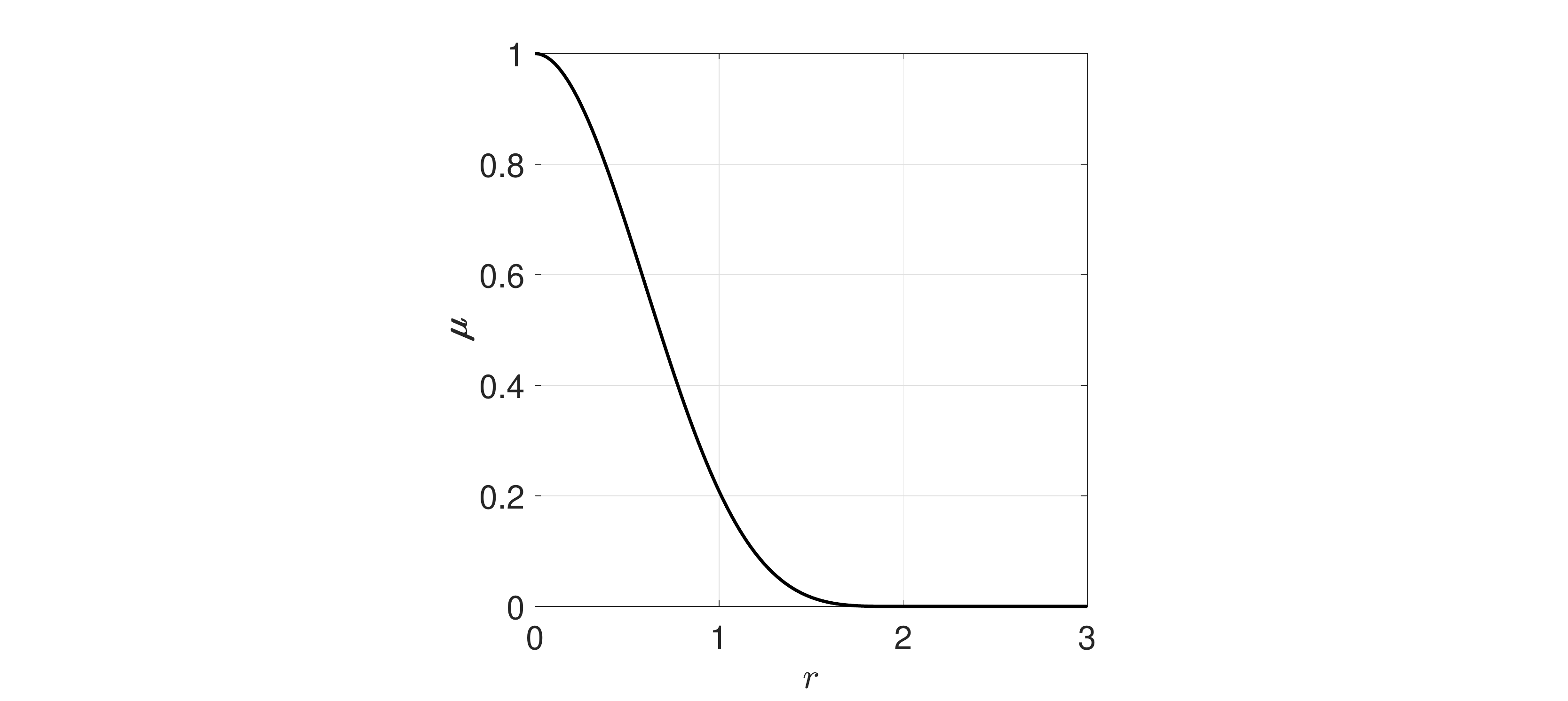}}
  \caption{Gaspari–Cohn correlation function $\bm{\mu}(r)$ used in the localization.}
\label{fig:gcf}
\end{figure}

The local-correlation function $\bm{\mu}$ in the localization equation \eqref{eq:loc} is defined as the Gaspari–Cohn (GC) function~\citep{gaspari1999construction}, given by
\begin{equation}
    [\bm{\mu}]_{ij}=\begin{cases}1-\frac{5}{3}r^2+\frac{5}{8}r^3+\frac{1}{2}r^4-\frac{1}{4}r^5
    ~&\text{for}~0\leq r < 1\\4-5r+\frac{5}{3}r^2+\frac{5}{8}r^3-\frac{1}{2}r^4+\frac{1}{12}r^5-\frac{2}{3r}~&\text{for}~1\leq r < 2   \\0~&\text{for}~r\geq 2
    \end{cases}
,
\end{equation}
where $r=|\boldsymbol{x}_i-\boldsymbol{x}_j|/(\sqrt{3}a/2)$, with $a$ taking the same value as in \eqref{eq:noise1}. A plot of $\bm{\mu}(r)$ is provided in figure~\ref{fig:gcf}.

\bibliographystyle{jfm}
\bibliography{jfm-instructions}

\end{document}